\documentclass[10pt,journal,compsoc]{IEEEtran}
\ifCLASSOPTIONcompsoc
  \usepackage[nocompress]{cite}
\else
  \usepackage{cite}
\fi
\usepackage{amsmath,amsfonts}
\usepackage{algorithmic}
\usepackage{algorithm}
\usepackage{array}
\usepackage[caption=false,font=normalsize,labelfont=sf,textfont=sf]{subfig}
\usepackage{textcomp}
\usepackage{stfloats}
\usepackage{url}
\usepackage{verbatim}
\usepackage{graphicx}
\usepackage{cite}

\newtheorem{theorem}{Theorem}
\newtheorem{proof}{Proof}

\usepackage{xcolor}
\usepackage{multirow}

\ifCLASSINFOpdf
\else
\fi
\begin{document}
%
\title{{\sffamily Arena}: A Learning-based Synchronization Scheme
for Hierarchical Federated Learning -- Technical Report}
%
%
%
%

\author{
Tianyu~Qi,~Yufeng~Zhan,~Peng~Li,~\IEEEmembership{Senior~Member,~IEEE},~Jingcai~Guo,\\~and~Yuanqing~Xia,~\IEEEmembership{Senior~Member,~IEEE}
\IEEEcompsocitemizethanks{
\IEEEcompsocthanksitem T.~Qi,~Y.~Zhan, and Y.~Xia are with the School of Automation, Beijing Institute of Technology, Beijing, China. 
E-mail: qitianyu@bit.edu.cn,~yu-feng.zhan@bit.edu.cn,~xia\_yuanqing@bit.edu.cn.
\IEEEcompsocthanksitem P. Li is with the School of Computer Science and Engineering, The University of Aizu, Aizuwakamatsu, Japan. 
E-mail: pengli@u-aizu.ac.jp.
\IEEEcompsocthanksitem J.~Guo is with the Department of Computing, The Hong Kong Polytechnic University, Hong Kong SAR, China. 
E-mail: jc-jingcai.guo@polyu.edu.hk.
}
\thanks{The conference version of this paper is ``Hwamei: A Learning-based Aggregation Framework for Hierarchical Federated Learning System," in Proc. of IEEE ICDCS, 2023.}}

\IEEEtitleabstractindextext{%
\begin{abstract}
Federated learning (FL) enables collaborative model training among distributed devices without data sharing, but existing FL suffers from poor scalability because of global model synchronization. To address this issue, hierarchical federated learning (HFL) has been recently proposed to let edge servers aggregate models of devices in proximity, while synchronizing via the cloud periodically. However, a critical open challenge about how to make a good synchronization scheme (when devices and edges should be synchronized) is still unsolved. Devices are heterogeneous in computing and communication capability, and their data could be non-IID. No existing work can well synchronize various roles (\textit{e.g.}, devices and edges) in HFL to guarantee high learning efficiency and accuracy. In this paper, we propose a learning-based synchronization scheme for HFL systems. By collecting data such as edge models, CPU usage, communication time, \textit{etc}., we design a deep reinforcement learning-based approach to decide the frequencies of cloud aggregation and edge aggregation, respectively. The proposed scheme well considers device heterogeneity, non-IID data and device mobility, to maximize the training model accuracy while minimizing the energy overhead.
%
Meanwhile, the convergence bound of the proposed synchronization scheme has been analyzed.
%
And we build an HFL testbed and conduct the experiments with real data obtained from Raspberry Pi and Alibaba Cloud. Extensive experiments under various settings are conducted to confirm the effectiveness of \textit{Arena}.
\end{abstract}

}

\maketitle

\IEEEdisplaynontitleabstractindextext

%
\IEEEpeerreviewmaketitle

\section{Introduction}
\IEEEPARstart{F}{ederated} learning (FL)~\cite{FedAvg,lai2021oort} has been proposed as a new learning paradigm that can exploit the knowledge hidden in distributed data while preserving data privacy. FL's basic idea is to let devices train their own models independently, and then share them with the cloud, which aggregates these models and sends aggregation results back to devices for knowledge sharing. Although we have witnessed some beautiful theoretical works and successful deployment of some small-scale FL testbeds \cite{DBLP:conf/micro/TianLSW022,AutoFL,Favor}, it is still hard to see large-scale FL deployment. The main reason is the poor scalability of existing FL. When the number of devices involved in FL increases to a certain level, the FL efficiency degrades seriously or even fails\cite{TiFL,FedAT}. 

\begin{figure}[!tb]
    \centering
    \includegraphics[width=0.6\linewidth]{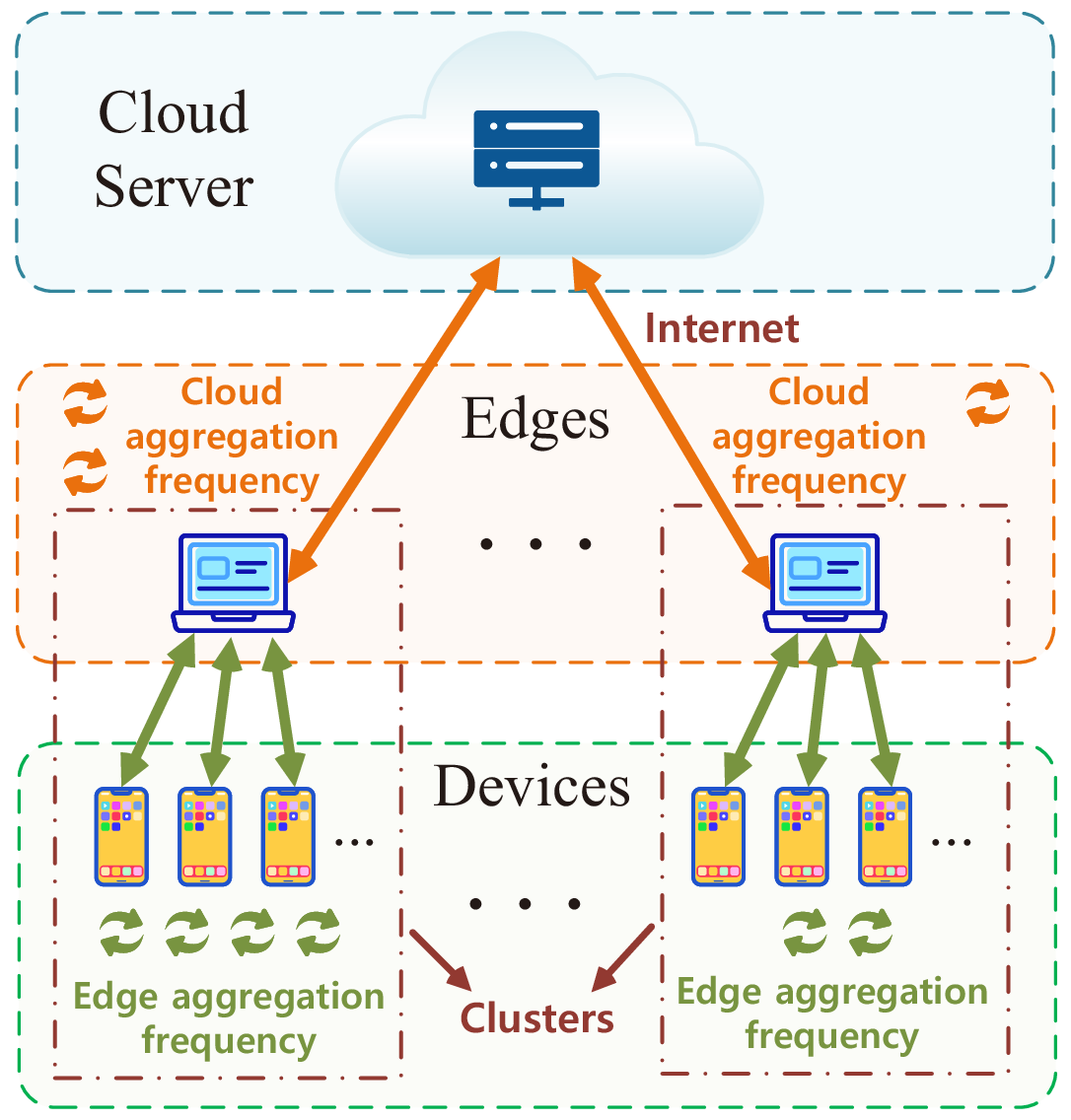}
    \caption{A synchronization scheme based on HFL framework}
    \label{fig:HFL}
    \vspace{-5mm}
\end{figure}

There are many research efforts on how to improve FL's scalability. As shown in Fig.~\ref{fig:HFL}, one of the most promising and practical solutions is hierarchical federated learning (HFL)~\cite{HFL}, which divides devices into several clusters and employs an edge server in each cluster to aggregate models of devices in proximity. The cloud conducts global model aggregation among edge servers. Note that HFL could have more layers, but we constrain our discussion to a 2-layer structure in this paper because it is typical and easy to implement in practice.
The effectiveness of HFL has been shown by existing work\cite{Share,RFLHA,FedCS,lim2021decentralized}. 
%
Deng~et~al.~\cite{Share} explores the optimization of training by altering the topological structure, yet it fails to comprehensively consider system heterogeneity. Authors in~\cite{RFLHA} propose a method that combines synchronous and asynchronous aggregation and~\cite{FedCS} utilize the DQN algorithm to determine the staleness of asynchronous aggregation. But the asynchronous approach exacerbates communication overhead.
However, there is still an open challenge about how to make a good synchronization scheme, \textit{i.e.}, how many training rounds are needed by devices before local edge aggregation (\textit{i.e.}, edge aggregation frequency) and when should we run the global cloud aggregation (\textit{i.e.}, cloud aggregation frequency),
to fully play the power of HFL. 
%
%
Our experiments show that there could be a big performance gap (\textit{e.g.}, 31.2\% in model accuracy increases and 36.4\% in energy consumption of devices) when applying different synchronization schemes. 

\IEEEpubidadjcol

However, designing an efficient synchronization scheme is much more difficult in HFL than in traditional FL. Compared to the single layer structure of traditional FL, HFL has more layers, and each layer may adopt different synchronization schemes, which significantly enlarges the search space of solutions. Moreover, hardware heterogeneity, non-IID data and device mobility further increase the difficulty. Hardware heterogeneity stems from the fact that devices have different hardware and network conditions, thus leading to different training speeds~\cite{SmartPC}. 
%
Meanwhile, even with the same hardware, devices may share different portions of CPU and memory resources to HFL. This is common because multiple applications run on our mobile devices, and HFL usually has low running priority to avoid negative influence on user experiences. 
For example, when games are running on smartphones, they prioritize resources to ensure users experiences.
%
Therefore, it is important to consider this hardware heterogeneity and dynamic computing and communication capability in synchronization scheme design. Second, it has been shown that non-IID data has a significant influence to traditional FL \cite{AutoFL,DBLP:conf/micro/TianLSW022,lai2021oort,wang2020towards}, but it is still unclear how it affects the HFL. We find that although non-IID data bring additional challenges, we can minimize their influences by carefully designing the synchronization scheme. Finally, device mobility plays a critical role in HFL. Some devices may join or leave HFL at any time. The synchronization scheme needs to be flexible to such a dynamic caused by device mobility. Note that all three factors are not independent and should be jointly considered in the synchronization scheme design.

To tackle the above challenges, in this paper, we present \textit{Arena}, a le\underline{Ar}ning-bas\underline{e}d sy\underline{n}chroniz\underline{a}tion scheme of HFL to intelligently control the aggregation frequencies of the cloud and edges. We first explore the time and power consumption of machine learning training tasks under different CPU usage caused by different interference programs. 
%
We propose a deep reinforcement learning~(DRL) approach based on proximal policy optimization~(PPO), aiming to guide the design of the synchronization scheme. In each cloud aggregation, \textit{Arena} collects edge models, device energy consumption, training time, \textit{etc}. DRL states and the reward function are sophisticatedly designed by fully considering energy consumption and training performance.
%
%
We construct an HFL testbed using Raspberry Pi devices and Alibaba Cloud, and conduct various experiments to verify the superiority of \textit{Arena} by comparing it with the state-of-the-arts. 
%
Compared to our previous conference version~\cite{qi2023hwamei}, we optimize the design of our algorithm. We enhance the profiling module and completely reengineer the action selection and reward function of DRL in this study, which improves the training efficiency of the agent and the final convergence performance.
Meanwhile, we theoretically analyzed the convergence guarantee of this aggregation scheme.
The results demonstrate that \textit{Arena} can significantly improve model accuracy and reduce device energy consumption. We highlight our major contributions as follows.

\begin{itemize}
    \item We propose \textit{Arena}, an intelligent HFL synchronization scheme, which decides the aggregation frequencies of cloud and edge. \textit{Arena} can co-optimize the model performance and training efficiency. 
    We provide a detailed convergence proof of the proposed synchronization scheme.
    \item We develop an HFL testbed with Raspberry Pi and Alibaba Cloud. The model training time and energy consumption under different CPU usage and edge-to-cloud communication time are collected, and these data are applied to decide aggregation frequencies.
    \item To verify the effectiveness of \textit{Arena}, MNIST and Cifar-10 are used to conduct the experiment. 
    Comparing with the state-of-the-arts, \textit{Arena} greatly improves the model accuracy by 31.2\% while reducing the energy consumption by 36.4\%.
\end{itemize}

The remainder of this paper is organized as follows. We present the background and motivation in Section~\ref{Second}. In Section~\ref{Third}, we describe the design of \textit{Arena}. We conduct extensive experiments in Section~\ref{Fourth}, and the related work is reviewed in Section~\ref{Fifth}. Finally, we conclude this paper in Section~\ref{Sixth}.

\section{Background and Motivation} \label{Second}

In this section, we will briefly introduce the fundamentals of hierarchical federated learning and conduct preliminary experiments to motivate the design of \textit{Arena}.

\subsection{Hierarchical Federated Learning}
HFL introduces edges between devices and the cloud.
%
Each edge is connected to some devices to form a cluster. The cloud communicates with edges, and edges communicate with the devices in their clusters. 
Assume there are $M$ edges in an HFL system.
%
Let $j$ denote an edge which connects $N_j$ device, then $\sum_{j=1}^M N_j=N$. 

In each cloud aggregation process, the cloud first sends the global model to the edges and devices. Then each device~$i$ updates the local model. After implementing~$\gamma_1$ epochs of local training, each device submits the local model~$w_i$ to the corresponding edge. When the edge receives the models of all devices in the cluster, it starts to aggregate the new model as
\begin{equation}
    w_j^e=\sum _{i=1}^{N_j} \frac{\left | \mathcal{D}_i \right | w_i}{\sum _{i=1}^{N_j}\left | \mathcal{D}_i \right |}, 
\end{equation}
where $w_j^e$ is the model of edge $j$ after aggregation. Then the edge will send $w_j^e$ back to the corresponding devices for continuous training. When the number of edge aggregation reaches $\gamma_2$, all edges submit their models to the cloud for cloud aggregation as
\begin{equation}
    w=\sum _{j=1}^{M}\frac{\left | \mathcal{D}_j \right | w_j^e}{\sum _{j=1}^{M}\left | \mathcal{D}_j \right |}, 
\end{equation}
where $\mathcal{D}_j$ is the dataset collection of all devices in the cluster of edge $j$. 
%
After $K$ rounds of cloud aggregation, we can get the final target global model. 

We consider a problem of minimizing empirical risk when training the neural network model, as
\begin{eqnarray} 
\min _{w} f(w)=\min _{w} \frac{1}{\left |  \mathcal{D}\right |} \sum _{(x,y)\in \mathcal{D}}f(w,x,y),
\end{eqnarray}
where $f(\cdot)$ indicates the loss function.
For SGD algorithm, the model parameters evolve as 
\begin{eqnarray} 
w(k)=w(k-1)-\eta \tilde{\nabla} f\left(w(k-1)\right).
\end{eqnarray}
After each round of cloud aggregation, the model updates can be summarized according to the HFL framework as~\cite{HFL}
\begin{equation}\label{old_lemma_1}
\begin{aligned}
    &w(k+1)=w(k)- \\
    &\eta \sum_{j \in[M]} \frac{N_{j}}{N} \frac{1}{N_{j}} \sum_{\alpha=0}^{\gamma_{2,k}^j-1} \sum_{i \in [N_{j}]} \sum_{\beta=0}^{\gamma_{1,k}^j-1} \tilde{\nabla} f_i\left(w_i\left(k, \alpha, \beta \right)\right),
\end{aligned}
\end{equation}
where $f_i(\cdot)$ is the loss function of device~$i$, and $w_i\left(k, \alpha, \beta \right)$ represents the model of device $i$ during the $k$-th round of cloud communication, which has undergone $\beta$ local training iterations and $\alpha$ edge aggregations.

\begin{figure}[!tb]
\centering
\subfloat[]{\includegraphics[width=3.0in]{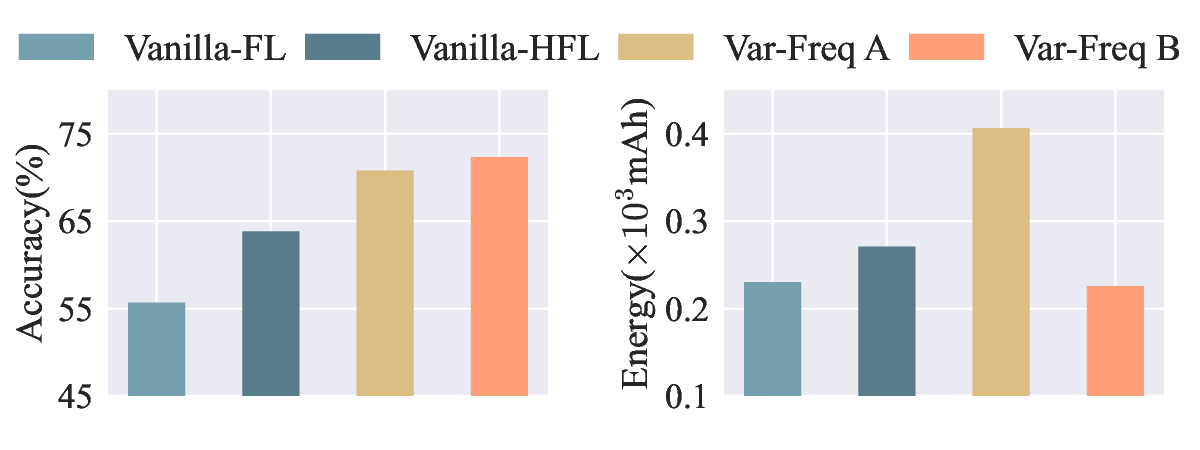}%
\label{fig:motivation1}}
\quad
\subfloat[]{\includegraphics[width=3.0in]{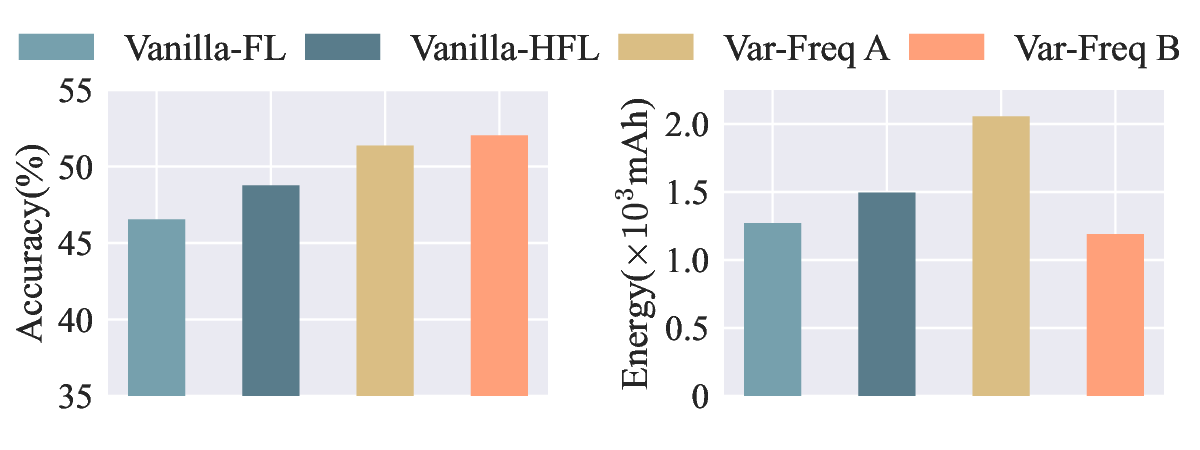}%
\label{fig:motivation2}}
\caption{Accuracy and energy within different frameworks. (a) The results of MNIST. (b) The results of Cifar-10.}
\label{fig:motivation}
\end{figure}

\subsection{How Synchronization Scheme Affects HFL}\label{sec2_B}

We have built an HFL testbed consisting of an Alibaba cloud server, 5 laptops as edge servers, and 50 Raspberry Pi as devices. In the vanilla HFL, all the devices have the same edge aggregation frequency and edges have the same cloud aggregation frequency. However, devices have different computing speeds and communication overhead, and thus some would become the stragglers. 
In the preliminary experiments, we design a new synchronization scheme by \underline{Var}ying the aggregation \underline{Freq}uency in the HFL framework called \textit{Var-Freq}.
%
Specifically, devices with similar training speeds are associated with edges by using a simple existing policy, \textit{e.g.}, K-cluster algorithm\cite{RFLHA}.
We present the details of clustering in Section~\ref{Third}-A. 
Besides clustering, we determine aggregation frequency. Each edge has its own cloud aggregation frequency, and all devices with similar training speeds under one edge share the same edge aggregation frequency.
%
We argue that properly adjusting the aggregation frequency during training can mitigate stragglers and improve model drift caused by statistical heterogeneity between edges. 
%

We train MNIST\cite{MNIST} and Cifar-10\cite{Cifar} by \textit{Vanilla-FL}, \textit{Vanilla-HFL}, \textit{Var-Freq A} and \textit{Var-Freq B}, with the same training time $3000$s for MNIST and $12000$s for Cifar-10. \textit{Vanilla-FL} is the traditional FL method that devices communicate with the cloud directly. It requires 1 hyperparameter to control the aggregation frequency. 
\textit{Vanilla-HFL} uses 2 hyperparameters $\gamma_1$ and $\gamma_2$ to control the edge and cloud aggregation frequency. 
Specifically, when we let $\gamma_2=1$, the \textit{Vanilla-HFL} method can be directly transformed into \textit{Vanilla-FL}. We control the same $\gamma_1 \cdot \gamma_2$ in \textit{Vanilla-FL} and \textit{Vanilla-HFL} to verify the excellence of the HFL framework under the same training times. Here we set $\gamma_1=20,\gamma_2=1$ for \textit{Vanilla-FL} and $\gamma_1=5,\gamma_2=4$ for \textit{Vanilla-HFL}. 
In \textit{Var-Freq A}, we first put devices under the edge according to the cluster. Then we appropriately increase the edge and cloud aggregation frequency of slower clusters, until all clusters have similar training times in each cloud communication round. 
However, unbalanced edge and cloud aggregation frequencies are prone to the stale model, which leads to model drift and waste of resources\cite{staleness}. For \textit{Var-Freq B}, we tune the parameters carefully on the basis of \textit{Var-Freq A}, \textit{i.e.}, appropriately reduce the aggregation frequency of fast devices with high energy consumption. We find optimized parameters of the aggregation frequency and get better results.

As shown in Fig.~\ref{fig:motivation}, we record each algorithm's termination accuracy and total energy consumption under different datasets. We can find that the accuracy of \textit{HFL} is higher than that of \textit{FedAvg}. This shows that \textit{HFL} performs better in model convergence, which is proved in \cite{HFLconverge}. At the same time, the accuracy of \textit{var-Freq A} is higher than that of \textit{HFL}, indicating that the method of clustering and changing the aggregation frequency is suitable for system heterogeneity. Since we simply increase the aggregation frequency of slow clusters, the energy consumption of \textit{var-Freq A} increases greatly. However, after adjusting the parameters, we can get the results of high accuracy and low energy consumption in \textit{var-Freq B}. This shows that a reasonable aggregation frequency can maximize model accuracy and energy efficiency but is challenging to find.

\subsection{System Dynamics}




\begin{figure}[!tb]
\centering
\subfloat[]{\includegraphics[width=3.0in]{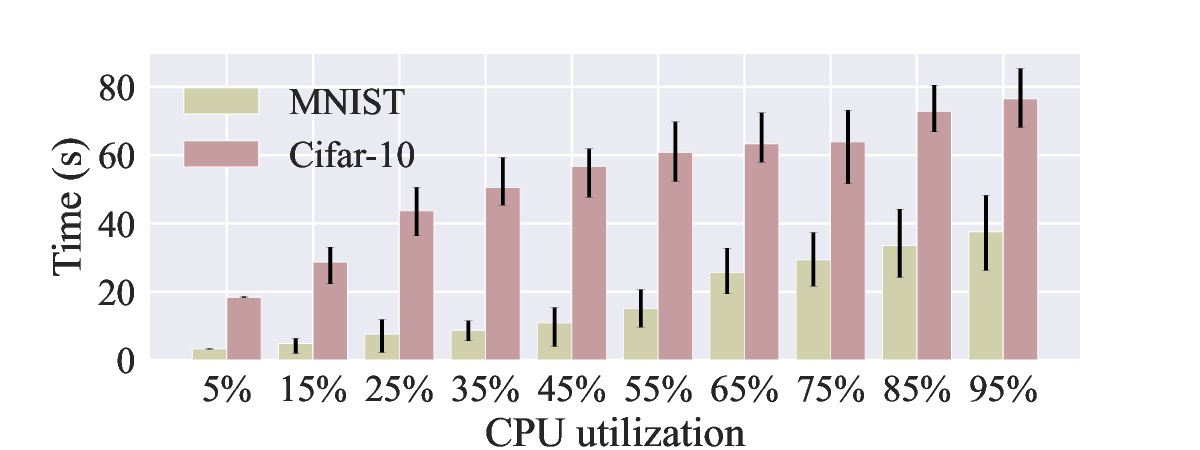}%
\label{fig:cpua}}
\quad
\subfloat[]{\includegraphics[width=3.0in]{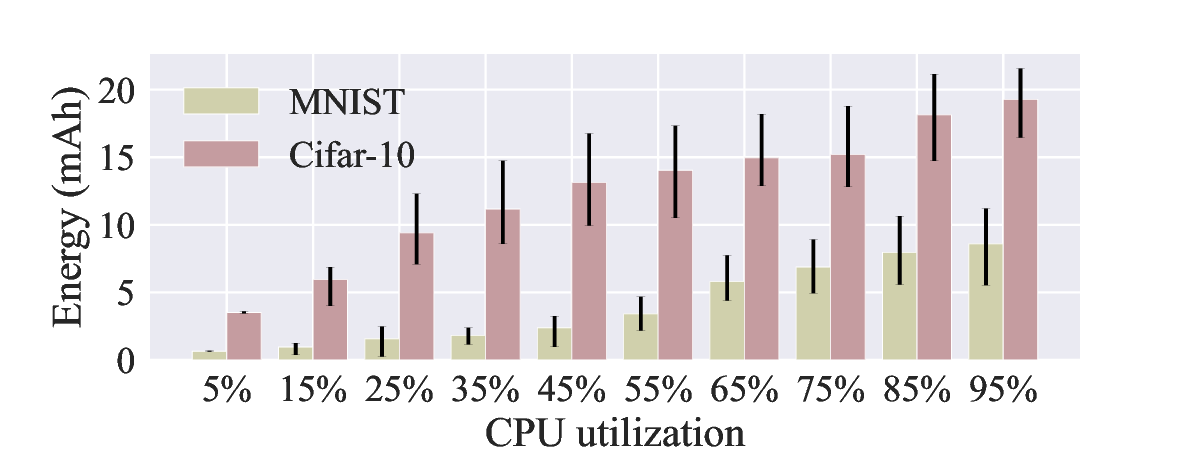}%
\label{fig:cpub}}
\caption{Training time and energy of Raspberry Pi with different CPU usage. (a) Single SGD training time at different CPU usage. (b) Single SGD training energy at different CPU usage.}
\label{fig:cpuusage}
\end{figure}

\textbf{Dynamic available CPU resources.} 
There are other programs running on the devices, which may interfere with the federated learning tasks. These programs also consume some CPU resources, and it is difficult to model such a resource competition because users' usage pattern is unpredictable.

To study this interference, we set the CPU of the Raspberry Pi to a conservative mode, which allows the CPU to allocate frequencies dynamically between 0.6GHz and 1.5GHz. Meanwhile, we use a CPU stress test tool, \textit{stress-ng}\cite{stressng}, to simulate interference programs. We change the available CPU resources of Raspberry Pi devices from 5\% to 95\%, and train MNIST and Cifar-10 under this setting. The Power Monitor\cite{powermonitor} is applied to measure the energy consumption of a single SGD process. In Fig.~\ref{fig:cpuusage}, we find that as the CPU usage increases, the time and energy consumption of the same training task increases. Under the same CPU setting, the training time and energy consumption fluctuate greatly. Mutual interference between programs and automatic frequency tuning of the CPU are the leading causes of fluctuations. The results above mean that the time and energy consumption during FL training is dynamic.

\begin{figure}[!tb]
    \centering
    \includegraphics[width=3.2  in]{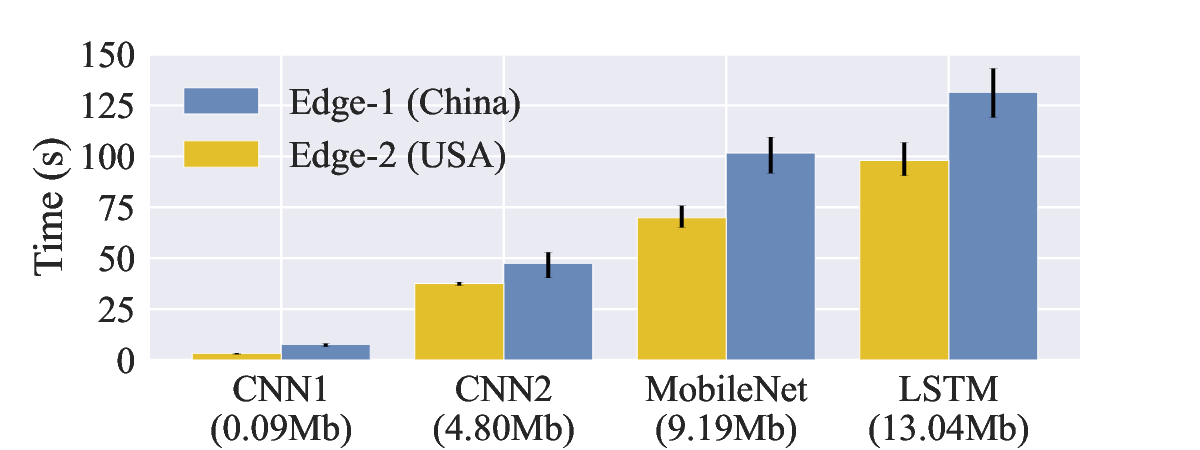}
    \caption{Edge-to-cloud communication time in different regions}
    \label{fig:edgetime}
\end{figure}

\textbf{Communication time at the edge.} 
Communication time can be divided into a device-to-edge part and an edge-to-cloud part. Since edges and devices are usually connected by high LAN, the device-to-edge communication overhead is at the millisecond level, which could be ignored. To verify the communication time from different edges to the cloud, we test the communication with local (China) and overseas (USA) edges to the same cloud. In Fig.~\ref{fig:edgetime}, we can find that cloud communication time grows as the model size increases. And since our cloud is deployed overseas, even for the same model, the edge communication time varies from one region to another.

Based on the above observations, we can conclude that: 1) changing the aggregation frequency of each edge and device after clustering can improve the training performance of federated learning; 
and 2) actual HFL systems could be highly dynamic, so a fixed synchronization setting can hardly work in practice. 
Inspired by these insights, we need a learning-based approach to guide the aggregation strategy in the heterogeneous and dynamic system, which achieves a good compromise between learning performance and energy consumption.

\section{Arena Design} \label{Third}

To optimize the model performance and minimize energy consumption in the presence of system heterogeneity, stochastic heterogeneity and fixed training time, we propose an DRL-based intelligent algorithm called \textit{Arena}. Specifically, an DRL agent deployed on the cloud server selects actions for the current state by accumulating rewards. In the HFL framework, the agent collects global information as the state without violating privacy and decides the number of edge and device training rounds in each round of cloud aggregation.

\begin{figure}[!tb]
    \centering
    \includegraphics[width=3.2  in]{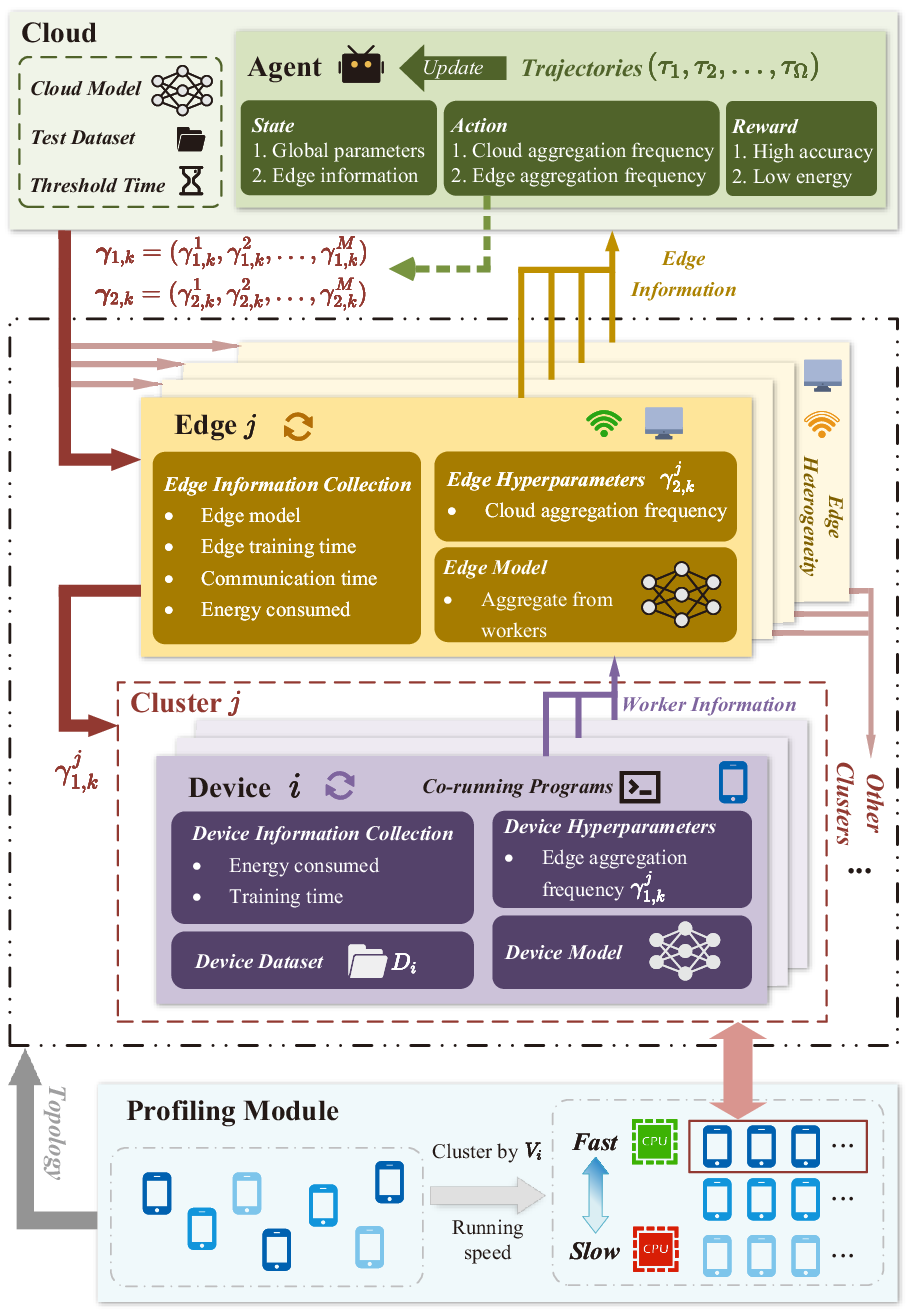}
    \caption{Overview of \textit{Arena}}
    \label{fig:Hwamei}
    \vspace{-5mm}
\end{figure}

Fig.~\ref{fig:Hwamei} shows the overview of \textit{Arena}. The agent is deployed in the cloud server. First, the devices are clustered according to computational speed and energy consumption by the profiling module. Devices with similar computing resources are assigned to the same edge. 
In each cloud aggregation, \textit{Arena} gets the global configuration parameters of the environment, including remaining time, energy consumption, and test accuracy. 
In addition, it collects edge information such as aggregation model, communication time, training time, \textit{etc}. 
\textit{Arena} dynamically assigns aggregation frequency to edges and devices for the captured state. 
Then the model is trained in the HFL framework according to the aggregation frequency, which expects to maximize model accuracy and energy efficiency. 
The trajectory is collected when the training time exceeds the threshold, and then the agent is updated by the trajectory.
%
%
%
Optimizing the system in this work by DRL involves two essential design requirements.

\textbf{High test accuracy.} Given a threshold time, we expect the system learns the optimal prediction model. 
As the models need to adapt quickly to new environments to maximize service quality
%
It is vital to learn models with higher accuracy in a limited time. To cope with dynamically changing environmental information, we need to model states and rewards. The design of the state takes into account all the heterogeneous information of the entire system as much as possible. 

\textbf{Low energy consumption.} Devices may have co-running programs. In Section~\ref{sec2_B}, we find that the time and energy consumption for HFL varies significantly among devices with different CPU usage. \textit{Arena} achieves low-energy consumption by carefully designing the cloud and edge aggregation frequencies.
%

%
The followings are the details of \textit{Arena}'s design.

\subsection{Profiling Module}

We find that the running time of the same device is not the same under different CPU usage in HFL. If the aggregation frequency is equally distributed, it will cause a straggler problem. With HFL's natural cluster framework, we can cluster devices and put them under the edge for training. We introduce the profiling module to initialize the topology to solve this challenge.

As shown in Fig.~\ref{fig:Hwamei}, all devices have their own characteristic $V_i$, $i \in \{1,2,\dots,N\}$. 
The $V_i$ contains 5 elements: configuration time $T_i^{pro}$, energy consumption $E_i^{pro}$, floating point operations per second $Fl_i^{pro}$, crystal frequency $Fr_i^{pro}$, and CPU utilization rate $Ut_i^{pro}$, \textit{i.e.}, $V_i=[T_i^{pro}\;E_i^{pro}\;Fl_i^{pro}\;Fr_i^{pro}\;Ut_i^{pro}]$.
%
At the ﬁrst step, the elements in $V_i$ of all available devices are set to 0. Then the profiling module lets all devices run the same number of epochs of the profiling task. The cloud server records the information of each device from the profiling task in this process and gets the final characteristic $V_i$. 
Next, these devices use $V_i$ to cluster according to the $K$-cluster algorithm, which minimizes the mean square error and balances the cluster size. 
In this study, we employ the $AFK-MC^2$ algorithm\cite{AFK-MC2}, which enhances the initial seed selection of k-means++ and enables fast clustering for large-scale systems.
%
Finally, all devices are assigned to the appropriate side, and the computing resources of devices on each side are guaranteed to be similar. For large-scale systems, we divide edges and devices into multiple groups by region(\textit{e.g.}, factories, companies), then we cluster devices under each group. If new devices join, the profiling module can also periodically re-cluster. This will not have any effect on the input and output dimensions of the DRL agent.

The devices in each edge cluster can greatly overcome the straggler effect. Some previous work also applies clustering to eliminate the straggler challenge\cite{FedAT,RFLHA}. But most of them are asynchronous algorithms, which depend on the design of staleness treatment. Our synchronization algorithm allows the agent to dynamically allocate aggregation frequencies, thus avoiding the impact of stale models.

\subsection{State}

\begin{figure}[!tb]
    \centering
    \includegraphics[width=1.0\linewidth]{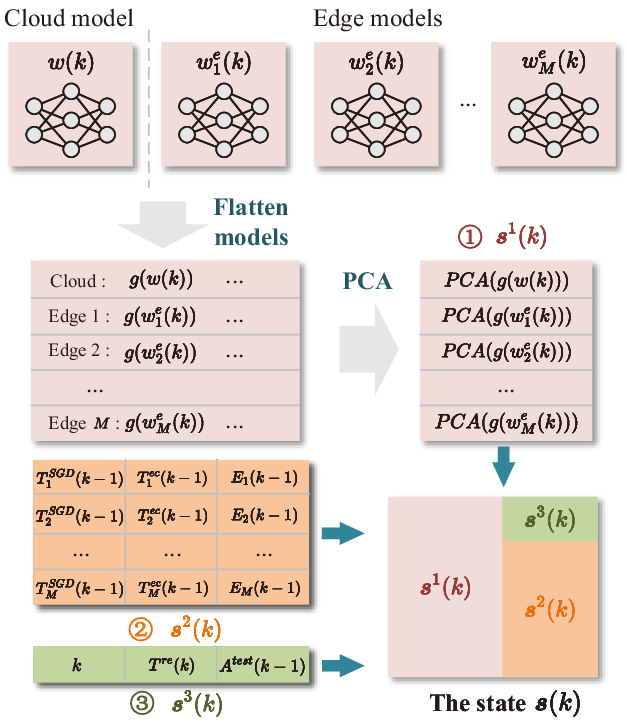}
    \caption{Composition of the state}
    \label{fig:state}
    \vspace{-4mm}
\end{figure}

Based on the principle of not violating devices' privacy, we get as many parameters as possible and compress them as the state. The state input is a two-dimensional array divided into two parts regarding acquisition sources: global parameters and edge information. Global parameters include the cloud model information and some configuration parameters. Edge information includes edge models and important information about devices in the cluster. As shown in Fig.~\ref{fig:state}, we split these parameters into multiple subsets for interpretation.

First, we collect the models of cloud and edges, \textit{i.e.}, $(w(k),w_1^e(k),w_2^e(k),\dots,w_M^e(k))$. The model parameters can reflect the data heterogeneity of different devices. 
Since our agent is deployed in the cloud, so there is no need to obtain device models, thus protecting user privacy to a certain extent. 
%
%
The final result shows that the agent is good enough to analyze the heterogeneity among edges by the designed state.

We flatten the model parameters and arrange them in order to make it easier for the agent to learn the difference among edges. However, using all model parameters as states will result in a large state space, makings it challenging to train the DRL agent with a space of millions of dimensions. We apply principle component analysis (PCA) to compress the models. The experiments in \cite{Favor} prove that the PCA compressed model can reflect the data heterogeneity among devices. The part of the state representing models can be expressed as
\begin{equation}
    \boldsymbol{s}^1(k) = PCA\{[g(w(k))^T \ g(w^e_1(k))^T\ \dots\ g(w^e_M(k))^T]^T\},
\end{equation}
where $g(\cdot)$ indicates the operation of flattening the model. Speciﬁcally, in the first cloud aggregation, we train HFL by a fixed aggregation frequency, then we compute the PCA loading vectors of different principal components based on the model. Because the principal components of models have enough information to identify the data distribution after the first cloud aggregation. In the latter cloud aggregation, the PCA loading vectors are reused to transform the models without fitting the PCA model again.

In addition to the models of edges, edge $j$ needs to provide the following information as part of the state
\begin{equation}
    \boldsymbol{h}_j(k)=[T^{SGD}_j(k-1)\ T_j^{ec}(k-1)\ E_j(k-1)]^T,
\end{equation}
where $T^{SGD}_j(k-1)$ denotes the local SGD time of the slowest device under the edge $j$ in the last cloud aggregation. It reflects the straggler's training information of the devices under edge $j$. $T_j^{ec}(k-1)$ is the communication time from the edge $j$ to the cloud in the last cloud aggregation. It represents the heterogeneity of edge communication. $E_j(k-1)$ is the energy that the devices consumed in the last cloud aggregation under edge $j$. All the information at the edge constitutes the second part of the state
\begin{equation}
    \boldsymbol{s}^2(k) = [\boldsymbol{h}_1(k)\ \boldsymbol{h}_2(k)\ \dots\ \boldsymbol{h}_M(k)]^T.
\end{equation}

There are also three crucial global parameters in the state: the current round of cloud aggregation $k$, remaining training time $T^{re}(k)$, and the test accuracy $A^{test}(k-1)$ of the last cloud aggregation $k-1$. They reflect the progress of the training and constitute the third part of the state
\begin{equation}
    \boldsymbol{s}^3(k) = [k\ T^{re}(k)\ A^{test}(k-1)].
\end{equation}

All the state information is finally concatenated together to form a two-dimensional matrix as
\begin{equation}
    \boldsymbol{s}(k) = cat\{(\boldsymbol{s}_1(k),cat\{(\boldsymbol{s}_3(k),\boldsymbol{s}_2(k)),dim=0\}),dim=1\}.
\end{equation}
The concatenation of the states can be seen more clearly in Fig.~\ref{fig:state}. The shape of the state $\boldsymbol{s}(k)$ is $(M+1)\times (n_{PCA}+3)$. The $n_{PCA}$ is the principal component of the PCA module, and we set it to $6$ in our system. In the actor-network, we use CNN to extract the features of the state further.

\subsection{Action}

Actions in DRL represent the tunable control knobs of a system. At the beginning of each round $k$, the agent needs to decide the training hyperparameters. Our agent applies multiple continuous actions to control training, that is, the edge aggregation frequency $\boldsymbol{\gamma}_{1,k}=\{\gamma^1_{1,k},\gamma_{1,k}^2,\dots,\gamma_{1,k}^M\}$ and the cloud aggregation frequency $\boldsymbol{\gamma}_{2,k}=\{\gamma_{2,k}^1,\gamma_{2,k}^2,\dots,\gamma_{2,k}^M\}$. Because we put devices with similar computing resources under the same edge, devices under the same edge giving the same aggregation frequency will not have stragglers. Clustering allows us not to decide the aggregation frequency for each device, which reduces the dimensionality of actions.

The $\boldsymbol{\gamma}_{1,k}$ and $\boldsymbol{\gamma}_{2,k}$ contain a total of $2M$ parameters. The actor-network in the agent has $4M$ output dimensions, and every two outputs form a distribution as expectation and variance. We can get $2M$ continuous values by sampling from $2M$ Gaussian distributions. After rounding and removing negative values, we can get $\gamma_{1,k}^j,\gamma_{2,k}^j\in \mathbb{N},j\in \{1,2,\dots,M\}$ finally.

\subsection{Reward}

Reward tracks the optimization objective in DRL. After $k$-th cloud aggregation, our system will give a reward $r(k)$ to the agent. We aim to reduce total energy consumption with high model prediction accuracy. Based on the principle of keeping the design of rewards as simple as possible, we set the reward after $k$-th cloud aggregation as
\begin{equation}
    r(k)=\Upsilon^{A^{test}(k)}-\Upsilon^{A^{test}(k-1)}-\epsilon E(k).
\end{equation}
The reward can be divided into two parts. The first part $\Upsilon^{A^{test}(k)}-\Upsilon^{A^{test}(k-1)}$ represents the increased accuracy of this training round. 
We discover that, for machine learning training tasks, the accuracy improvement tends to slow down as it approaches convergence. In order to let the DRL agent can capture the small model improvement near the end of the HFL process, we use a constant $\Upsilon$ which is set to 64, to optimize the design of the reward.
The second part, $E(k)$, is the total power consumption of all devices in this round, \textit{i.e.}, $E(k)=\sum_{j=1}^M{E_j(k)}$. And $\epsilon \in (0,1)$ is the weight used to balance accuracy and energy consumption. When the accuracy of this cloud aggregation round improves significantly, the agent will get a higher reward. When unnecessary devices consume too much energy, the reward value becomes lower.

The agent is trained to maximize the expectation of the cumulative discounted reward given by
\begin{equation}
    R(k)=\sum_{k=1}^{K^{*}} \xi^{k-1} (\Upsilon^{A^{test}(k)}-\Upsilon^{A^{test}(k-1)}-\epsilon E(k)),
\end{equation}
where $\xi\in(0,1]$ is a factor discounting future rewards. The trajectory length $K^*$ is not fixed. It is determined by the threshold time $T$. Applying this reward to guide the agent can train a better model with less energy consumption.
\begin{algorithm}[!tb]
    \caption{\textit{Arena}'s Training Process}\label{algorithm}
    \begin{algorithmic}[1]
        \STATE Initialize the topology by profiling module;
        \STATE Initialize threshold time $T$, remaining time $T^{re}(0)=T$, global model $w(0)$, round of cloud aggregations $k=0$;
        \STATE Train once cloud aggregation by given aggregation frequencies, get $w(1)$, $w_j^e(1)$, and record $T^{use}(1)$;
        \STATE Train PCA module by $w(1)$ and $w_j^e(1)$;
        \STATE Update $T^{re}_{init}=T^{re}(1)-T^{use}(1)$; $k$++;
        \FOR{$1$ \TO $\Omega$ } 
            \WHILE {\TRUE}
                \STATE Observe state $\boldsymbol{s}(k)$;
                \STATE Choose actions $\boldsymbol{a}(k)$, that is $\boldsymbol{\gamma}_{1,k}$ and $\boldsymbol{\gamma}_{2,k}$;
                \STATE Train HFL by $\boldsymbol{\gamma}_{1,k}$ and $\boldsymbol{\gamma}_{2,k}$, record $T^{use}(k)$;
                \STATE Get reward $r(k)$ and $\boldsymbol{s}({k+1})$, update $T^{re}(k)=T^{re}(k-1)-T^{use}(k)$; 
                \STATE Push $(\boldsymbol{s}(k),\boldsymbol{a}(k),r(k),\boldsymbol{s}({k+1}))$ to agent memory; 
                \STATE $k$++;
                \IF{$T^{re}(k)<0$}
                    \STATE Set $k=1$, $T^{re}(k)=T^{re}_{init}$; 
                    \STATE \textbf{break}
                \ENDIF
            \ENDWHILE
            \STATE Update the agent and clear agent memory;
        \ENDFOR
    \end{algorithmic}
\end{algorithm}

\subsection{Workflows}

Algorithm~\ref{algorithm} shows the detailed algorithm for initial configuration and DRL training in \textit{Arena}, following the steps below:

\textbf{Step 1.} Initialize the topology by profiling module (Line 1). Set threshold time $T$ and remaining time $T^{re}(0)=T$. Initialize the HFL model as $w(0)$ (Line 2);

\textbf{Step 2.} In the first cloud aggregation, train HFL by given aggregation frequency, then train the PCA module by models of cloud and edges, \textit{i.e.}, $w(1)$ and $w^e_j(1)$. Record $T^{use}(k)=\max\{T_j^{SGD}(k)+T_j^{ec}(k)\},j\in \{1,2,\dots,M\}$ and update the remaining time $T^{re}(k)$ (Line 3-5);

\textbf{Step 3.} The agent observe current state $\boldsymbol{s}(k)$ and decide the action $\boldsymbol{a}(k)$, that is $\boldsymbol{\gamma}_{1,k}$ and $\boldsymbol{\gamma}_{2,k}$. After training HFL by actions, we can get the reward $r(k)$, new state $\boldsymbol{s}({k+1})$, and new $T^{re}(k)$ of this round. Then push $(\boldsymbol{s}(k),\boldsymbol{a}(k),r(k),\boldsymbol{s}({k+1}))$ to agent memory (Line 12);

\textbf{Step 4.} Repeat step 3 until $T^{re}(k)<0$. We end this HFL training and reset $T^{re}(k)$ and $k$ (Line 14-17);

\textbf{Step 5.} Using state-action-reward in the memory pool to optimize the agent and clear the agent memory (Line 19). Repeat steps 3-5 for $\Omega$ times, which is the number of DRL episodes (Line 6-20).

The additional overhead introduced by \textit{Arena} is minimum. Because the edge model required to construct the state is necessary for HFL aggregation. The PCA is the most computational step in the workflows, and we put it on the cloud instead of the devices. \textit{Arena} also has high scalability. We can dynamically add devices to the system and re-clustering. If the number of edges remains the same, the state dimension does not change either. The agent can adjust actions by observing the changing training time and model parameters.

\subsection{Enhancement}

The agent obtains a large number of trajectories interactively during the training process. In particular, our use of multidimensional continuous actions leads to a great difference between the two distributions, so the importance sampling effect will not be satisfactory. 
%
We apply proximal policy optimization~\cite{schulman2017proximal} to optimize the DRL agent as
\begin{equation}
    \begin{aligned}
    J_{}^{\theta^{\prime }}&(\theta)  \approx \sum_{(\boldsymbol{s}(t), \boldsymbol{a}(t))} \min \left(\frac{p_\theta\left(\boldsymbol{a}(t) \mid \boldsymbol{s}(t)\right)}{p_{\theta^{\prime }}\left(\boldsymbol{a}(t) \mid \boldsymbol{s}(t)\right)} A^{\theta^{\prime }}\left(\boldsymbol{s}(t), \boldsymbol{a}(t)\right),\right. \\
    &\left.\operatorname{clip}\left(\frac{p_\theta\left(\boldsymbol{a}(t) \mid \boldsymbol{s}(t)\right)}{p_{\theta^{\prime}}\left(\boldsymbol{a}(t) \mid \boldsymbol{s}(t)\right)}, 1-\varepsilon, 1+\varepsilon\right) A^{\theta^{\prime}}\left(\boldsymbol{s}(t), \boldsymbol{a}(t)\right)\right),
    \end{aligned}
\end{equation}
where $\operatorname{clip}(\cdot)$ indicates that the first term is too large or too small and will be constrained to $1+\varepsilon$ and $1-\varepsilon$. The advantage of this is that the difference between $p_\theta\left(\boldsymbol{a}(t) \mid \boldsymbol{s}(t)\right)$ and $p_{\theta^{\prime}}\left(\boldsymbol{a}(t) \mid \boldsymbol{s}(t)\right)$ is not too much.

Besides, due to the dynamics of the system, the rewards for a large amount of data have high variance. This causes the agent to converge to a poor solution. We add generalized advantage estimation (GAE) to optimize the estimated variance of the reward. It estimates a more instructive value function by reward shaping as
\begin{equation}
    \hat{A}(t)^{\mathrm{GAE}}=\sum_{l=0}^{\infty}(\xi \lambda)^l \tilde{r}\left(\boldsymbol{s}({t+l}), \boldsymbol{a}(t), \boldsymbol{s}({t+l+1})\right),
\end{equation}
where $\tilde{r}\left(\boldsymbol{s}({t+l}), \boldsymbol{a}(t), \boldsymbol{s}({t+l+1})\right)=\xi V(\boldsymbol{s}({t+l+1}))-V(\boldsymbol{s}({t+l}))$, and $V(\boldsymbol{s}(t))+r(t)$ is the value obtained by the critic network when the state is $\boldsymbol{s}(t)$. The $\lambda\in (0,1]$ is a smoothing parameter used for reducing the variance. This approach looks at a longer future to get higher rewards. 

Regarding the actions in DRL, we have also implemented the following optimizations. We search for the solution with the minimum distance within all solution spaces for the generated continuous action values, \textit{i.e.}, $\min \left \| \widetilde{\boldsymbol{a}(t)} - \boldsymbol{a}(t)\right \|_2^2 $, where $\widetilde{\boldsymbol{a}(t)}$ represents all possible combinations of solutions for frequencies $\boldsymbol{\gamma}_{1,k}$ and $\boldsymbol{\gamma}_{2,k}$. This design allows for a more precise determination of the actual action for each round rather than rounding.

\subsection{Convergence Analysis of Arena}
\textit{Arena} applies the synchronization scheme by varying the aggregation frequency in the HFL framework. Here, we provide the convergence bound for this aggregation scheme. We assume the loss function satisfies following two assumptions.

\textbf{Assumption 1}: The loss function is $L$-smooth and the Lipschitz constant $L>0$, \textit{i.e.}, $\|\nabla f(x)-\nabla f(y)\| \leq L\|x-y\|$.

\textbf{Assumption 2}: The estimated stochastic gradient is unbiased and the variance of stochastic gradients on each device is bounded, \textit{i.e.}, $\mathbb{E}\left[\left\|\tilde{\nabla} f_i(w_i)-\nabla f(w)\right\|^2 \mid w\right] \leq \sigma^2$.


 With Assumption~1 and~2, we can get the relationship between $w(k)$ and $w(k+1)$ after one cloud aggregation is
\begin{equation} \label{lemma2}
\begin{aligned} 
&\mathbb{E} [f\left(w(k+1)\right)]-\mathbb{E} [f\left(w(k)\right)] \leq \frac{L}{2} \mathbb{E}\left\|{w}({k+1})-w(k)\right\|^2 \\
& + \mathbb{E}\left\langle\nabla f\left(w(k)\right),{w}({k+1})-w(k)\right\rangle.
\end{aligned}
\end{equation}



\begin{theorem}\label{theorem-1}
    The convergence bound of one cloud aggregation is
\begin{eqnarray}
\begin{aligned} \label{theorem1}
&\mathbb{E} [f\left(w({k+1})\right)]-\mathbb{E} [f\left(w(k)\right)] \\
\leq& \frac{L^2 \eta^3}{4} \widetilde{\boldsymbol{\gamma}}_1 \widetilde{\boldsymbol{\gamma}_2}\left(\left(\widetilde{\boldsymbol{\gamma}_1}-1\right)+\frac{M}{N} \widetilde{\boldsymbol{\gamma}_1}\left(\widetilde{\boldsymbol{\gamma}_2}-1\right)\right) \sigma^2 \\
&+\frac{L \eta^2}{2} \frac{1}{N} \widetilde{\boldsymbol{\gamma}_1} \widetilde{\boldsymbol{\gamma}_2} \sigma^2 -\frac{\eta}{2} \widetilde{\boldsymbol{\gamma}_1} \widetilde{\boldsymbol{\gamma}_2} \mathbb{E}\left\|\nabla f\left(w(k)\right)\right\|^2,
\end{aligned}
\end{eqnarray}
where $\widetilde{\boldsymbol{\gamma}_1}$ and $\widetilde{\boldsymbol{\gamma}_2}$ are the maximum values of $\boldsymbol{\gamma}_{1,k}$ and $\boldsymbol{\gamma}_{2,k}$, respectively. 
\end{theorem} 

\begin{proof}
We separately prove the convergence domain for the two parts on the right-hand side of \eqref{lemma2}. We first prove the range of the latter part.

\textbf{Part 1}: After taking the expectation, we get \eqref{25}. According to $2\langle\boldsymbol{a}, \boldsymbol{b}\rangle=\|\boldsymbol{a}\|^2+\|\boldsymbol{b}\|^2-\|\boldsymbol{a}-\boldsymbol{b}\|^2$, we achieve \eqref{26}. Regarding the third term in \eqref{26}, by combining $\mathbb{E}[\|x\|^2]=\|\mathbb{E} [x]\|^2+\operatorname{Var}(x)^2$, we can deduce the \eqref{ABC}, where $j_i$ represents the cluster $j$ which the device $i$ belongs to. 
And $N_{j_i}$ and $\gamma_{1,k}^{j_i}$ respectively represent the number of devices and the aggregation frequency of this cluster. We can bound the three terms of \eqref{ABC} as \eqref{A}, \eqref{B} and \eqref{C}, where $\widetilde{\boldsymbol{\gamma_1}}$ is the max value from $\{\gamma_{1,k}^1,\gamma_{1,k}^2,\dots,\gamma_{1,k}^M\}$. 
Then \eqref{ABC} can be written as \eqref{D1D2}. Substituting into \eqref{25} and \eqref{26}, the limits of $D_1$ is
\begin{small}
\begin{equation}
\begin{aligned}
    &\frac{1}{N} \sum_{j \in[M]} \sum_{i \in [N_{j}]} \sum_{t_2=0}^{\gamma_{2,k}^j-1} \sum_{t_1=0}^{\gamma_{1,k}^j-1} D_1 \\
    \leq & \frac{1}{N} \sum_{j \in[M]} \sum_{i \in [N_{j}]} \frac{\gamma_{1,k}^j(\gamma_{1,k}^j-1)}{2} \sum_{\alpha=0}^{\gamma_{2,k}^j-1} \sum_{\beta=0}^{\gamma_{1,k}^j-1} \mathbb{E}\left\|\nabla f\left(w({k, \alpha, \beta})\right)\right\|^2 \\
    +&\frac{1}{N} \sum_{j \in[M]} \sum_{i \in [N_{j}]} \frac{1}{N_{j_i}} \sum_{i^\prime \in [N_{j_i}]} \sum_{\alpha=0}^{\gamma_{2,k}^{j_i}-1} \sum_{\beta=0}^{\gamma_{1,k}^{j_i}-1} \mathbb{E}\left\|\nabla f\left(w_{i^\prime}({k, \alpha, \beta})\right)\right\|^2 \\
    &\times \frac{(\gamma_{2,k}^j-1)\gamma_{2,k}^j}{2}\widetilde{\boldsymbol{\gamma_1}}^2\\
    =& \frac{1}{N} \sum_{j \in[M]} \sum_{i \in [N_{j}]} \left( \frac{\gamma_{1,k}^j(\gamma_{1,k}^j-1)}{2} + \frac{(\gamma_{2,k}^j-1)\gamma_{2,k}^j}{2}\widetilde{\boldsymbol{\gamma_1}}^2 \right) \times \\
    &\sum_{\alpha=0}^{\gamma_{2,k}^j-1} \sum_{\beta=0}^{\gamma_{1,k}^j-1} \mathbb{E}\left\|\nabla f\left(w({k, \alpha, \beta})\right)\right\|^2.
\end{aligned}
\end{equation}
\end{small}
And the limits of $D_2$ is
\begin{equation}
\begin{aligned}
    & \frac{1}{N} \sum_{j \in[M]} \sum_{i \in [N_{j}]} \sum_{t_2=0}^{\gamma_{2,k}^j-1} \sum_{t_1=0}^{\gamma_{1,k}^j-1} D_2 = \frac{1}{N} \sum_{j \in[M]} \sum_{i \in [N_{j}]} \cdot \\
    &\left( \frac{\gamma_{2,k}^j\gamma_{1,k}^j(\gamma_{1,k}^j-1)}{2}\sigma^2 + \frac{\gamma_{1,k}^j\gamma_{2,k}^j(\gamma_{2,k}^j-1)}{2} \cdot \frac{\widetilde{\boldsymbol{\gamma_1}}}{N_{j_i}}\sigma^2 \right) \\
    \leq& \frac{\widetilde{\boldsymbol{\gamma_2}} \widetilde{\boldsymbol{\gamma_1}}}{2}\left(\left(\widetilde{\boldsymbol{\gamma_1}}-1\right)+\frac{M}{N} \widetilde{\boldsymbol{\gamma_1}}\left(\widetilde{\boldsymbol{\gamma_2}}-1\right)\right) \sigma^2.
\end{aligned}
\end{equation}
Combining \eqref{25} and \eqref{26}, we can obtain the final result of Part 1 as shown in \eqref{part1}.

\begin{figure*}[!bp]
\begin{small}
\begin{equation}\label{25}
\begin{aligned}
    \mathbb{E}\left\langle\nabla f\left(w(k)\right), {w}({k+1})-w(k)\right\rangle &=-\mathbb{E}\left\langle\nabla f\left(w(k)\right), \eta \sum_{j \in[M]} \frac{1}{N}  \sum_{\alpha=0}^{\gamma_{2,k}^j-1} \sum_{i \in [N_{j}]} \sum_{\beta=0}^{\gamma_{1,k}^j-1} \tilde{\nabla} f_i\left(w_i({k, \alpha, \beta})\right)\right\rangle \\
    &=-\eta\sum_{j \in[M]} \sum_{i \in [N_{j}]} \frac{1}{N} \sum_{t_2=0}^{\gamma_{2,k}^j-1} \sum_{t_1=0}^{\gamma_{1,k}^j-1} \mathbb{E}\left\langle\nabla f\left(w(k)\right),\nabla f\left(w_i({k,t_2,t_1})\right) \right\rangle.
\end{aligned}
\end{equation}
\end{small}
\end{figure*}
\begin{figure*}[!bp]
\begin{small}
\begin{equation}\label{26}
\begin{aligned}
    &-\mathbb{E}\left\langle\nabla f\left(w(k)\right), \nabla f\left(w_i({k, t_2, t_1})\right)\right\rangle=-\frac{1}{2} \mathbb{E}\left\|\nabla f\left(w(k)\right)\right\|^2- \frac{1}{2} \mathbb{E}\left\|\nabla f\left(w_i({k, t_2, t_1})\right)\right\|^2+\frac{1}{2} \mathbb{E}\left\|\nabla f\left(w(k)\right)-\nabla f\left(w_i({k, t_2, t_1})\right)\right\|^2.
\end{aligned}
\end{equation}
\end{small}
\end{figure*}
\begin{figure*}[!bp]
\begin{small}
\begin{equation}\label{ABC}
\begin{aligned}
&\mathbb{E}\left\|\nabla f\left(w(k)\right)-\nabla f\left(w_i({k, t_2, t_1})\right)\right\|^2 =L^2 \eta^2 \mathbb{E}\left\|\sum_{\beta=0}^{t_1-1} \tilde{\nabla} f_i\left(w_i({k, t_2, \beta})\right)+\sum_{\alpha=0}^{t_2-1} \sum_{i^\prime \in [N_{j_i}]} \frac{1}{N_{j_i}} \sum_{\beta=0}^{\gamma_{1,k}^{j_i}-1} \tilde{\nabla} f_{i^\prime} \left(w_{i^\prime} ({k, \alpha, \beta})\right)\right\|^2 \\
&= \left(  \underbrace{\mathbb{E}\left\|\sum_{\beta=0}^{t_1-1} \nabla f\left(w_i({k, t_2, \beta})\right)+\sum_{\alpha=0}^{t_2-1} \sum_{i^\prime \in [N_{j_i}]} \frac{1}{N_{j_i}} \sum_{\beta=0}^{\gamma_{1,k}^{j_i}-1} \nabla f\left(w_i^\prime ({k, \alpha, \beta})\right)\right\|^2}_A+\underbrace{\sum_{\beta=0}^{t_1-1} \mathbb{E}\left\|\left[\tilde{\nabla} f_i\left(w_i({k, t_2, \beta})\right)-\nabla f\left(w_i({k, t_2, \beta})\right)\right]\right\|^2}_B \right. \\
& \left. +\underbrace{\sum_{\alpha=0}^{t_2-1} \mathbb{E}\left\|\sum_{i^\prime \in [N_{j_i}]} \frac{1}{N_{j_i}} \sum_{\beta=0}^{\gamma_{1,k}^{j_i}-1} \tilde{\nabla} f_{i^\prime}\left(w_{i^\prime}({k, \alpha, \beta})\right)-\sum_{\beta=0}^{\gamma_{1,k}^{j_i}-1} \nabla f\left(w_{i^\prime}({k, \alpha, \beta})\right) \right\|^2}_{C} \right) \cdot L^2 \eta^2
.
\end{aligned}
\end{equation}
\end{small}
\end{figure*}
\begin{figure*}[!bp]
\begin{small}
\begin{eqnarray}\label{A}
\begin{aligned}
A \leq t_1 \sum_{\beta=0}^{t_1-1} \mathbb{E}\left\|\nabla f\left(w({k, t_2, \beta})\right)\right\|^2+ \frac{1}{N_{j_i}} t_2 \widetilde{\boldsymbol{\gamma}_1} \sum_{i^\prime \in [N_{j_i}]} \sum_{\alpha=0}^{t_2-1} \sum_{\beta=0}^{\gamma_{1,k}^{j_i}-1} \mathbb{E}\left\|\nabla f\left(w_{i^\prime}({k, \alpha, \beta})\right)\right\|^2.
\end{aligned}
\end{eqnarray}
\end{small}
\end{figure*}
\begin{figure*}[!bp]
\begin{eqnarray}\label{B}
B=\sum_{\beta=0}^{t_1-1} \mathbb{E}\left\|\left[\tilde{\nabla} f_i\left(w_i({k, t_2, \beta})\right)-\nabla f\left(w_i({k, t_2, \beta})\right)\right]\right\|^2 \leq t_1 \sigma^2.
\end{eqnarray}
\end{figure*}

\begin{figure*}[!htbp]
\begin{eqnarray}\label{C}
\begin{aligned}
C & =\sum_{\alpha=0}^{t_2-1} \sum_{i^\prime \in [N_{j_i}]} \frac{1}{(N_{j_i})^2} \mathbb{E}\left\| \sum_{\beta=0}^{\gamma_{1,k}^{j_i}-1} \tilde{\nabla} f_{i^\prime}\left(w_{i^\prime}({k, \alpha, \beta})\right) - \sum_{\beta=0}^{\gamma_{1,k}^{j_i}-1} \nabla f\left(w_{i^\prime}({k, \alpha, \beta})\right) \right\|^2 \leq \frac{t_2\widetilde{\boldsymbol{\gamma_1}}\sigma^2}{N_{j_i}}.
\end{aligned}
\end{eqnarray}
\end{figure*}
\begin{figure*}[!htbp]
\begin{eqnarray}\label{D1D2}
\begin{aligned}
    & \mathbb{E}\left\|\nabla f\left(w(k)\right)-\nabla f\left(w_i({k, t_2, t_1})\right)\right\|^2 = L^2 \eta^2\left(A+B+C\right) \\
    & \leq L^2 \eta^2\{\underbrace{t_1 \sum_{\beta=0}^{t_1-1} \mathbb{E}\left\|\nabla f\left(w({k, t_2, \beta})\right)\right\|^2+\frac{t_2 \widetilde{\boldsymbol{\gamma_1}}}{N_{j_i}} \sum_{i^\prime \in [N_{j_i}]} \sum_{\alpha=0}^{t_2-1} \sum_{\beta=0}^{\gamma_{1,k}^{j_i}-1} \mathbb{E}\left\|\nabla f\left(w_{i^\prime}({k, \alpha, \beta})\right)\right\|^2}_{D_1}\} +L^2\eta^2\underbrace{(t_1\sigma^2+\frac{t_2\widetilde{\boldsymbol{\gamma_1}}}{N_{j_i}}\sigma^2)}_{D_2}.
\end{aligned}
\end{eqnarray}
\end{figure*}
\begin{figure*}[!htbp]
\begin{eqnarray}\label{part1}
\begin{aligned}
    & \mathbb{E}\left\langle\nabla f\left(w(k)\right), w({k+1})-w(k)\right\rangle \\
    \leq & -\frac{\eta}{2} \widetilde{\boldsymbol{\gamma_2}} \widetilde{\boldsymbol{\gamma_1}} \mathbb{E}\left\|\nabla f\left(w(k)\right)\right\|^2 -\frac{\eta}{2} \sum_{j \in[M]} \sum_{i \in [N_{j}]} \frac{1}{N} \sum_{t_2=0}^{\gamma_{2,k}^j-1} \sum_{t_1=0}^{\gamma_{1,k}^j-1} \mathbb{E}\left\|\nabla f\left(w_i({k, t_2, t_1})\right)\right\|^2 \\
    &+ \frac{L^2 \eta^3}{2} \left( \frac{1}{N} \sum_{j \in[M]} \sum_{i \in [N_{j}]} \left( \frac{\gamma_{1,k}^j(\gamma_{1,k}^j-1)}{2} + \frac{(\gamma_{2,k}^j-1)\gamma_{2,k}^j}{2}\widetilde{\boldsymbol{\gamma_1}}^2 \right) \sum_{\alpha=0}^{\gamma_{2,k}^j-1} \sum_{\beta=0}^{\gamma_{1,k}^j-1} \mathbb{E}\left\|\nabla f\left(w_i({k, \alpha, \beta})\right)\right\|^2 \right) \\
    &+ \frac{L^2 \eta^3}{2} \frac{\widetilde{\boldsymbol{\gamma_2}} \widetilde{\boldsymbol{\gamma_1}}}{2}\left(\left(\widetilde{\boldsymbol{\gamma_1}}-1\right)+\frac{M}{N} \widetilde{\boldsymbol{\gamma_1}}\left(\widetilde{\boldsymbol{\gamma_2}}-1\right)\right) \sigma^2 \\
    \leq &-\frac{\eta}{2} \cdot \frac{1}{N} \sum_{j \in[M]} \sum_{i \in [N_{j}]} \left( 1-L^2 \eta^2\left(\frac{\gamma_{1,k}^j(\gamma_{1,k}^j-1)}{2} + \frac{(\gamma_{2,k}^j-1)\gamma_{2,k}^j}{2}\widetilde{\boldsymbol{\gamma_1}}^2\right)\right)  \sum_{\alpha=0}^{\gamma_{2,k}^j-1} \sum_{\beta=0}^{\gamma_{1,k}^j-1} \mathbb{E}\left\|\nabla f\left(w_i({k, \alpha, \beta})\right)\right\|^2 \\
    &-\frac{\eta}{2} \widetilde{\boldsymbol{\gamma_1}} \widetilde{\boldsymbol{\gamma_2}} \mathbb{E}\left\|\nabla f\left(w(k)\right)\right\|^2 +\frac{L^2 \eta^3}{4} \widetilde{\boldsymbol{\gamma_1}} \widetilde{\boldsymbol{\gamma_2}}\left[\left(\widetilde{\boldsymbol{\gamma_1}}-1\right)+\frac{M}{N} \widetilde{\boldsymbol{\gamma_1}}\left(\widetilde{\boldsymbol{\gamma_2}}-1\right)\right] \sigma^2.
\end{aligned}
\end{eqnarray}
\end{figure*}
\begin{figure*}
\begin{equation}\label{part2}
\begin{aligned}
    & \mathbb{E}\left\|w({k+1})-w(k)\right\|^2 \\
    =& \eta^2 \mathbb{E}\left\|\frac{1}{N} \sum_{j \in[M]} \sum_{i \in [N_{j}]} \sum_{t_2=0}^{\gamma_{2,k}^j-1} \sum_{t_1=0}^{\gamma_{1,k}^j-1} \nabla f_i\left(w_i({k, t_2, t_1})\right)\right\|^2 +\\
    &\eta^2 \frac{1}{N^2} \sum_{j \in[M]} \sum_{i \in [N_{j}]} \sum_{t_2=0}^{\gamma_{2,k}^j-1} \mathbb{E}\left\|\sum_{t_1=0}^{\gamma_{1,k}^j-1} \tilde{\nabla} f_i\left(w_i({k, t_2, t_1})\right)-\sum_{t_1=0}^{\gamma_{1,k}^j-1} \nabla f\left(w_i({k, t_2, t_1})\right)\right\|^2 \\
    \leq& \eta^2 \frac{1}{N} \sum_{j \in[M]} \sum_{i \in [N_{j}]} \gamma_{2,k}^j \gamma_{1,k}^j \sum_{t_2=0}^{\gamma_{2,k}^j-1} \sum_{t_1=0}^{\gamma_{1,k}^j-1} \mathbb{E}\left\|\nabla f\left(w_i({k, t_2, t_1})\right)\right\|^2 + \eta^2 \frac{1}{N^2} \sum_{j \in[M]} \sum_{i \in [N_{j}]} \sum_{t_2=0}^{\gamma_{2,k}^j-1}\widetilde{\boldsymbol{\gamma_1}} \sigma^2. 
\end{aligned}
\end{equation}
\end{figure*}

\textbf{Part 2}: Regarding the convergence bound of the first part in \eqref{lemma2}, based on the assumption $\mathbb{E}[\|x\|^2]=\|\mathbb{E} [x]\|^2+\operatorname{Var}(x)^2$, the following proof can be derived as \eqref{part2}.

By simultaneously solving \eqref{part1} and \eqref{part2} from Part 1 and 2, along with \eqref{25} and \eqref{26}, we can obtain the following results, as
\begin{small}
\begin{equation}
\begin{aligned}
&\mathbb{E} f\left(w({k+1})\right)-\mathbb{E} f\left(w(k)\right) \leq  -\frac{\eta}{2} \widetilde{\boldsymbol{\gamma_1}} \widetilde{\boldsymbol{\gamma_2}} \mathbb{E}\left\|\nabla f\left(w(k)\right)\right\|^2 \\
&-\frac{\eta}{2} \cdot \frac{1}{N} \sum_{j \in[M]} \sum_{i \in [N_{j}]} \sum_{\alpha=0}^{\gamma_{2,k}^j-1} \sum_{\beta=0}^{\gamma_{1,k}^j-1} \mathbb{E}\left\|\nabla f\left(w_i({k, \alpha, \beta})\right)\right\|^2 \\
&\left(1- L^2 \eta^2\left(\frac{\gamma_{1,k}^j\left(\gamma_{1,k}^j-1\right)}{2}+\frac{\widetilde{\boldsymbol{\gamma_1}}^2 \gamma_{2,k}^j\left(\gamma_{2,k}^j-1\right)}{2}\right)-L \eta\gamma_{1,k}^j \gamma_{2,k}^j\right)\\
&+\frac{L^2 \eta^3}{4} \widetilde{\boldsymbol{\gamma_1}} \widetilde{\boldsymbol{\gamma_2}}\left(\left(\widetilde{\boldsymbol{\gamma_1}}-1\right)+\frac{M}{N} \widetilde{\boldsymbol{\gamma_1}}\left(\widetilde{\boldsymbol{\gamma_2}}-1\right)\right) \sigma^2+\frac{L \eta^2}{2N} \widetilde{\boldsymbol{\gamma_1}} \widetilde{\boldsymbol{\gamma_2}} \sigma^2.
\end{aligned}
\end{equation}
\end{small}
When $\eta$ is small enough, the system can converge if the following condition is satisfied
\begin{small}
\begin{equation}
\begin{aligned}
1-L^2 \eta^2\left(\frac{\gamma_{1,k}^j\left(\gamma_{1,k}^j-1\right)}{2}+\frac{\widetilde{\boldsymbol{\gamma_1}}^2 \gamma_{2,k}^j\left(\gamma_{2,k}^j-1\right)}{2}\right)-L \eta\gamma_{1,k}^j \gamma_{2,k}^j \geq 0.
\end{aligned}
\end{equation}
\end{small}
Then we can obtain the final convergence bound as \eqref{theorem1}. 
The convergence of the proposed synchronization
scheme in a cloud communication round has been demonstrated.
\end{proof}


\section{Evaluation} \label{Fourth}

In this section, we build a heterogeneous realistic HFL environment and train using the collected realistic data. We conduct experiments to prove the superiority of \textit{Arena} compared with advanced benchmarks.

\subsection{System Settings}

We build an actual HFL system similar to the preliminary experiments in Section~\ref{Second} via Raspberry Pi, laptops, and Alibaba cloud. Specifically, we use two standard datasets and corresponding models: 1) For MNIST, we use a CNN with 21,840 parameters composed of 2 convolutional layers and 2 fully connected layers. The dataset consists of 60,000 training samples and 10,000 test samples with 10 labels; 2) For Cifar-10, we use a CNN with 453,834 parameters composed of 3 convolutional layers and 3 fully connected layers. The dataset consists of 50,000 training samples and 10,000 test samples with 10 classes.

Heterogeneity includes data heterogeneity, computing heterogeneity, and communication heterogeneity. Our system has 50 devices and 5 edges. We set each device to have 2 classes with an equal amount of data (1,200 data per device for MNIST \& 1,000 data per device for Cifar-10), which indicates the system is non-IID. We set the CPU usage of the Raspberry Pi to 5 classes from 10\% to 50\% and 10 devices per class, which represents resource heterogeneity. Our cloud is deployed in Silicon Valley, USA. We test edge-to-cloud communications in Beijing, China, and Washington, D.C., USA, to simulate extreme edge communication heterogeneity. We assume 3 edges with random 30 devices in China and 2 edges with random 20 devices in the US, and edges and devices in the same area can form a cluster. This is much like a system consisting of different factories or companies. To conveniently train with real data, we record the edge communication time and apply it in the system training.

\begin{figure}[!tb]
\centering
\subfloat[]{\includegraphics[width=1.75in]{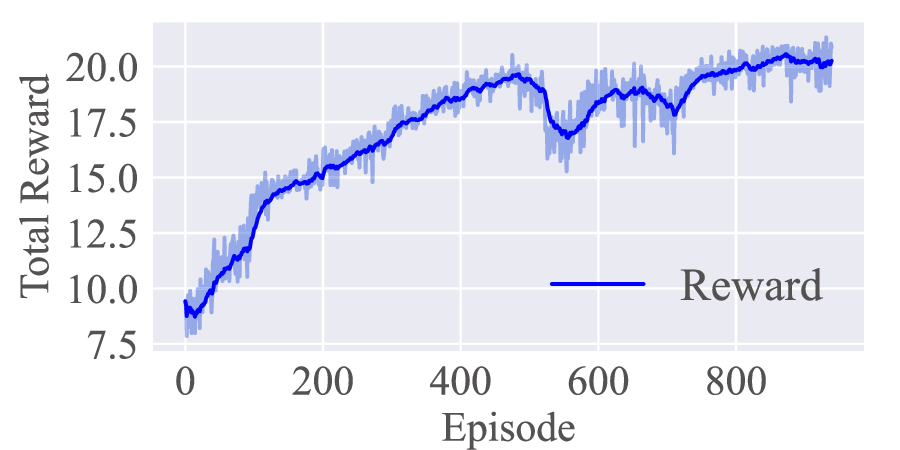}%
\label{fig:result1_MNIST_reward}}
\subfloat[]{\includegraphics[width=1.75in]{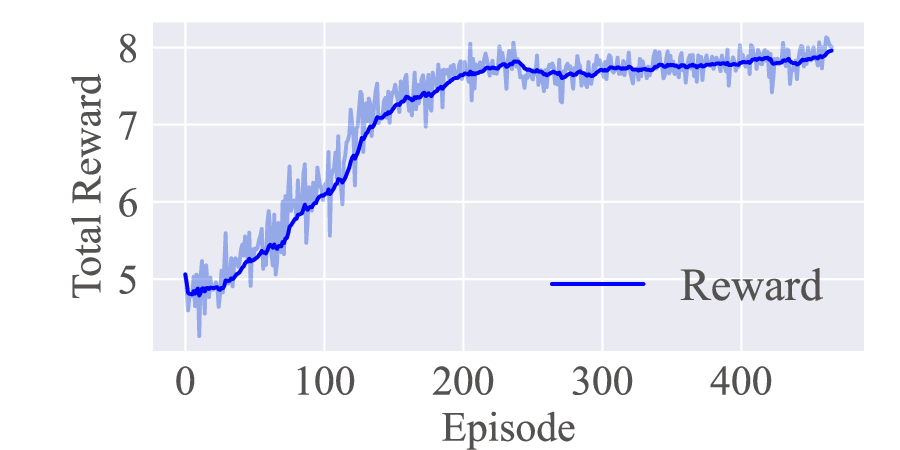}%
\label{fig:result1_Cifar_reward}}
\quad
\subfloat[]{\includegraphics[width=1.75in]{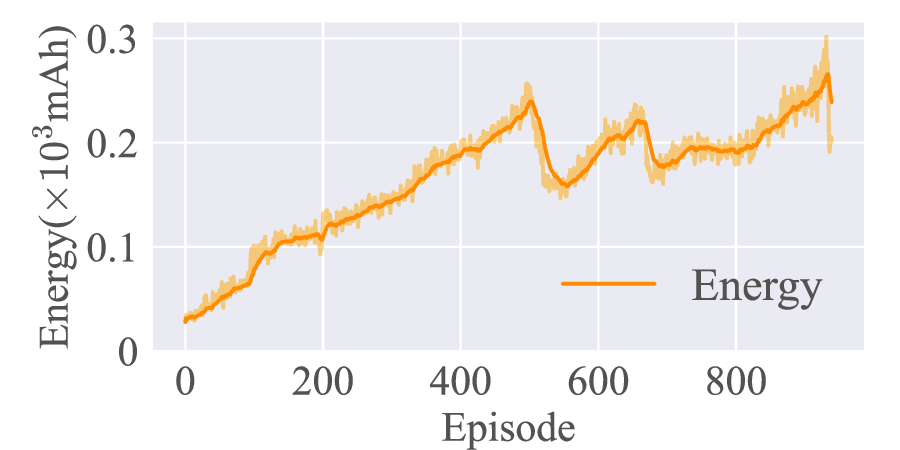}%
\label{fig:result1_MNIST_energy}}
\subfloat[]{\includegraphics[width=1.75in]{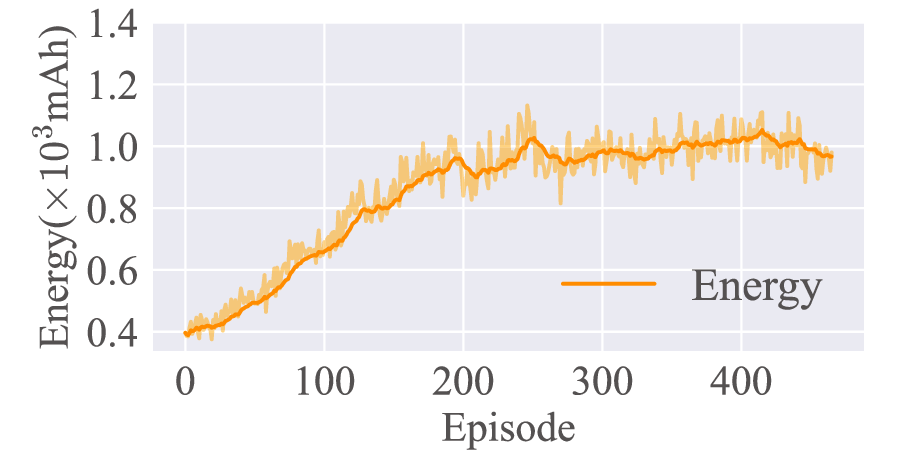}%
\label{fig:result1_Cifar_energy}}
\quad
\subfloat[]{\includegraphics[width=1.75in]{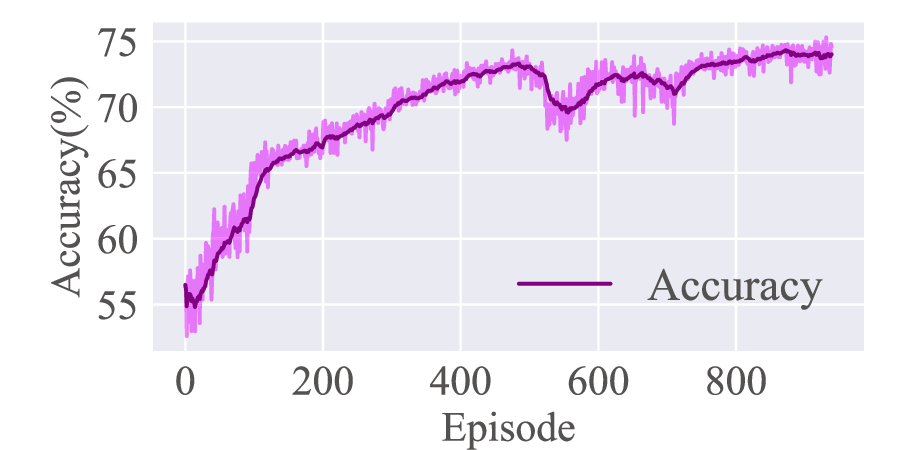}%
\label{fig:result1_MNIST_acc}}
\subfloat[]{\includegraphics[width=1.75in]{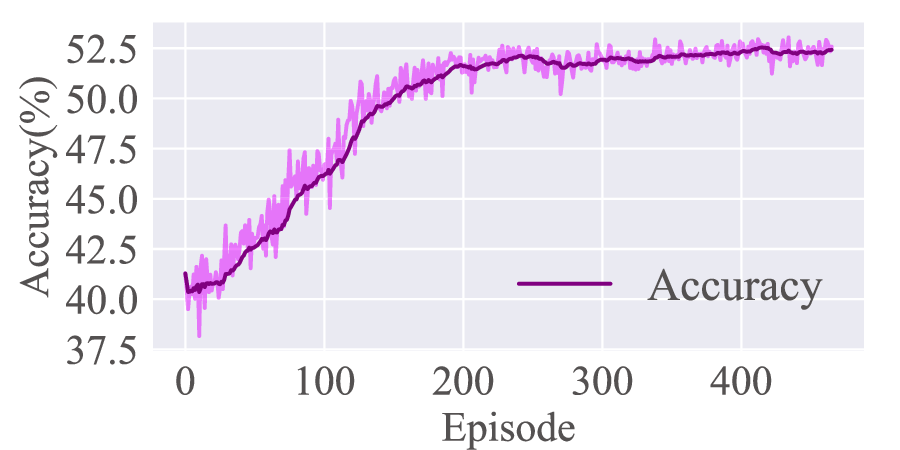}%
\label{fig:result1_Cifar_acc}}
\caption{Training the DRL agent of \textit{Arena}. (a)(b) Rewards from training MNIST and Cifar-10. (c)(d) Energy from training MNIST and Cifar-10. (e)(f) Accuracy from training MNIST and Cifar-10.}
\label{fig:DRLtrain}
\end{figure}

For performance comparison, we use the following algorithms as benchmarks:

\begin{itemize}
    \item \textit{Vanilla-FL}~\cite{FedAvg}: A baseline synchronous FL method. Devices communicate directly with the cloud without going through the edge. A certain percentage of devices are randomly selected for training in each round.
    \item \textit{Vanilla-HFL}\cite{HFL}: A baseline cloud-edge-device synchronous FL method. The system trains the models by fixed edge and cloud aggregation frequency at each cloud round.
    \item \textit{Favor}\cite{Favor}: A method based on FedAvg and DQN. The agent observes devices' models to select the appropriate devices to participate in the training at each round. 
    \item \textit{Share}\cite{Share}: A method that changes the topology of the HFL framework. The topology constructed with this algorithm solves the data distribution-aware communication cost minimization problem.
\end{itemize}


Additionally, our algorithm is designed and compared with our previous work, \textit{Hwamei}.

The training hyperparameters are set as follows. The SGD batch size is 32, and lr is 0.003 for MNIST and 0.01 for Cifar-10. We use 2 convolutional layers and 3 fully connected layers for the DRL network. The discount factor $\xi=0.9$ and the smoothing parameter $\lambda=0.9$ for GAE. For the clip function, we set $\varepsilon=0.2$. The reward weight $\epsilon=0.002$ for MNIST and $\epsilon=0.03$ for Cifar-10.

\subsection{Training the DRL Agent}

As shown in Fig.~\ref{fig:DRLtrain}, we train the DRL agent with the MNIST and Cifar-10 federated learning tasks. The threshold time we set $T=3000$s for MNIST and $T=12000$s for Cifar-10, and the number of DRL episodes for these two tasks is set to 1500 and 700, respectively. We can find the rewards of the tasks gradually converge as the episodes increase in Fig.~\ref{fig:result1_MNIST_reward} and \ref{fig:result1_Cifar_reward}. The average energy consumption of the devices will first increase and decrease later as shown in Fig.~\ref{fig:result1_MNIST_energy} and \ref{fig:result1_Cifar_energy}. Because after the aggregation frequency keeps increasing causing the accuracy to converge, our agent will find the solution that consumes the least amount of energy thanks to the design of rewards. And in Fig.~\ref{fig:result1_MNIST_acc} and \ref{fig:result1_Cifar_acc}, we observe that the final accuracy within the threshold time can reach 74.8\% for MNIST and 52.6\% for Cifar-10.

\begin{figure}[!tb]
\centering
\subfloat[]{\includegraphics[width=1.75in]{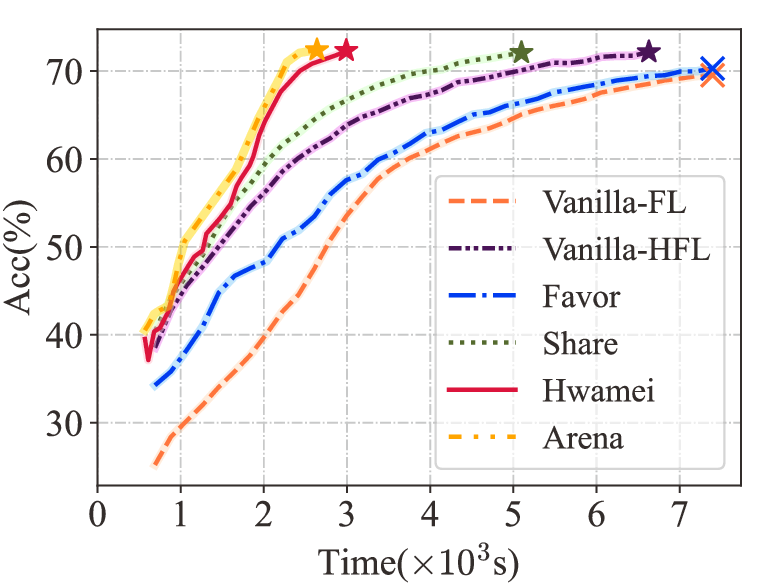}%
\label{fig:result2_MNIST}}
\subfloat[]{\includegraphics[width=1.75in]{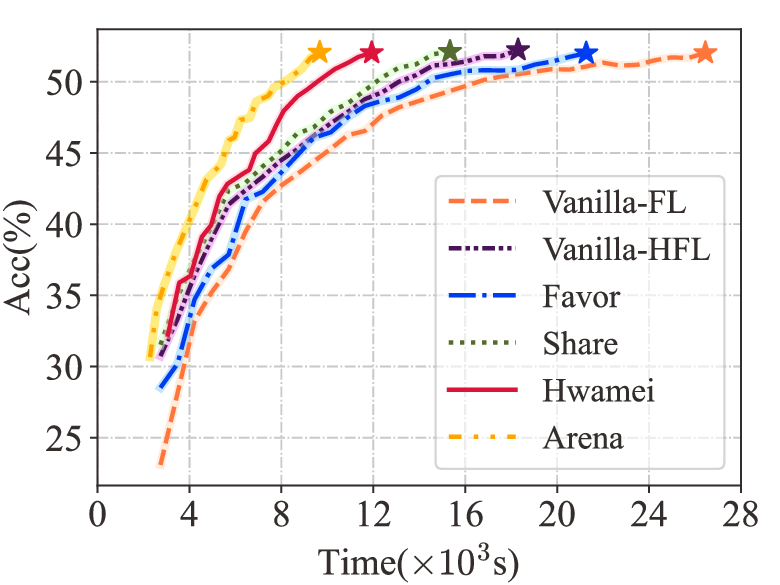}%
\label{fig:result2_Cifar}}
\caption{Relationship between time and accuracy of different FL methods. (a) Accuracy of training MNIST. (b) Accuracy of training Cifar-10.}
\label{fig:time_accuracy}
\end{figure}

To demonstrate the superiority of \textit{Arena}, we train the datasets to the same accuracy using benchmarks in the same environment. We test the benchmarks with several parameters and select the one that gives the best results. Fig.~\ref{fig:time_accuracy} shows the relationship between time and accuracy. Our algorithm shows excellent results in both datasets. In the early stages of training, \textit{Arena}'s accuracy is similar to that of \textit{Vanilla-HFL} and \textit{Share}, and the accuracy improves more significantly in the mid to late stages. In Fig.~\ref{fig:result2_MNIST} for MNIST, the accuracy of \textit{Vanilla-FL} and \textit{Favor} does not converge to 72\% for a long period. \textit{Vanilla-HFL} and \textit{Share} can achieve this accuracy but in times 6129s and 4597s, respectively. To reach this accuracy, \textit{Arena} saves 59.3\% and 41.2\% time for these two algorithms. The results for Cifar-10 in Fig.~\ref{fig:result2_Cifar} are similar to MNSIT. All four benchmarks achieve an accuracy of 52\%, but they all cost longer time than \textit{Arena}. To reach this accuracy, \textit{Arena} saves 63.3\% and 34.2\% compared to \textit{Vanilla-FL} and \textit{Share}, which takes the longest and shortest time in benchmarks, respectively. 
\textit{Arena} saves 14.1\% and 22.8\% of time compared to \textit{Hwamei} on two respective datasets.

\begin{figure}[!tb]
\centering
\subfloat[]{\includegraphics[width=3.5in]{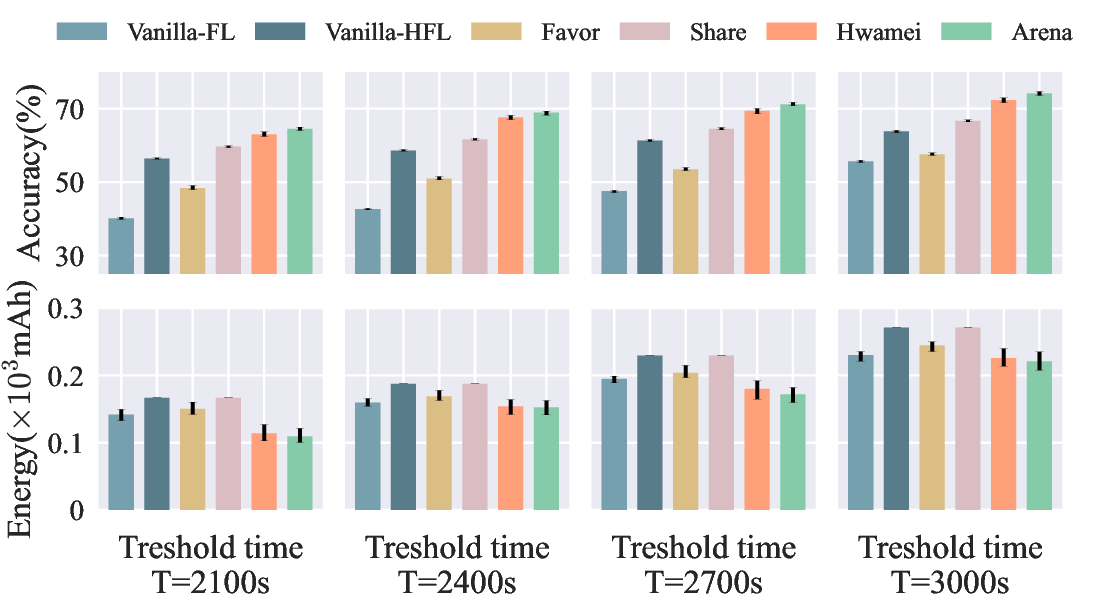}%
\label{fig:result3_MNIST}}
\quad
\subfloat[]{\includegraphics[width=3.5in]{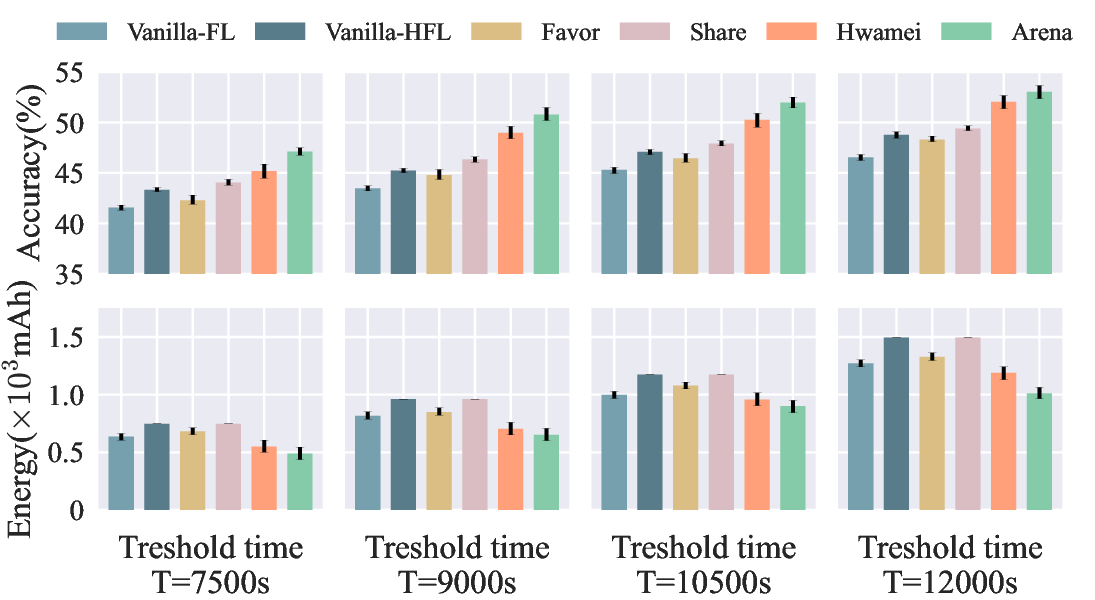}%
\label{fig:result3_Cifar}}
\caption{(a) Accuracy and energy of training MNIST within different threshold times. (b) Accuracy and energy of training Cifar-10 within different threshold times.}
\label{fig:result3}
\vspace{-4mm}
\end{figure}

\subsection{Test in Different Threshold Time}

In this part, we investigate the accuracy that each algorithm could achieve at different threshold times and the average energy consumption during training.

Fig.~\ref{fig:result3_MNIST} is the result for training MNIST. We select four threshold times and measure the maximum accuracy the algorithm could achieve within the threshold time and the average energy consumption during the period. For each algorithm, we test several times and get the results. As the threshold time increases, the accuracy of the different algorithms improves, and the energy overhead increase. This is in line with the expectations of FL methods. For the algorithms with different threshold times, we can find that the accuracy of algorithms based on hierarchical federated learning, \textit{i.e.}, \textit{Vanilla-HFL}, and \textit{Share}, is generally better than that of two-layer FL algorithms. However, \textit{Vanilla-FL} and \textit{Favor} consume less energy because they select some devices in each communication round. But our algorithm has the highest accuracy and lowest energy consumption at all times. At these four threshold times, \textit{Arena} improves accuracy by 30.7\% on average over \textit{Vanilla-FL}, which is the most insufficient accuracy in benchmarks. Compared to \textit{Share}, the highest accuracy in benchmarks, \textit{Arena} still improves accuracy by 10.2\% on average. In terms of energy consumption, the cost of \textit{Arena} is basically the same as that of \textit{Vanilla-FL}, which consumes the least in benchmarks. \textit{Arena} saves an average of 19.9\% of the cost per device compared with \textit{Vanilla-HFL}, which consumes the most energy in benchmarks.

The performance of \textit{Arena} in Cifar-10 is shown in Fig.~\ref{fig:result3_Cifar}. \textit{Arena} maintains the highest accuracy at different threshold times. Our method improves accuracy by 14.3\% over \textit{Vanilla-FL} and 7.1\% over \textit{Share} on average on the Cifar-10 task. \textit{Arena} still consumes the least amount of energy in this dataset. It can even save 22.4\% of energy consumption compared to \textit{Vanilla-HFL}. In summary, \textit{Arena} achieves that training of optimal prediction models with minimal energy overhead at any threshold time.

\textit{Arena} outperforms \textit{Hwamei} under different threshold time. \textit{Arena} has shown an average increase in accuracy of 2.38\% and 3.35\% compared to \textit{Hwamei}.

\renewcommand\arraystretch{1.2} 
\begin{table}[!tb]
\caption{Performance of cluster vs non-cluster on Arena}
\centering
\begin{tabular}{c|c|cc|cc}
\hline
\multirow{2}{*}{\textbf{}}         & \multirow{2}{*}{\textbf{Time}} & \multicolumn{2}{c|}{\textbf{Cluster}}                    & \multicolumn{2}{c}{\textbf{non-Cluster}}                 \\ \cline{3-6} 
                                   &                                & \multicolumn{1}{c|}{\textbf{Accuracy}} & \textbf{Energy} & \multicolumn{1}{c|}{\textbf{Accuracy}} & \textbf{Energy} \\ \hline
\multirow{4}{*}{\textbf{MNIST}}    & 2100s                          & \multicolumn{1}{c|}{64.5\%}              & 109mAh          & \multicolumn{1}{c|}{62.3\%}              & 128mAh          \\ \cline{2-6} 
                                   & 2400s                          & \multicolumn{1}{c|}{68.9\%}              & 152mAh          & \multicolumn{1}{c|}{66.9\%}              & 176mAh             \\ \cline{2-6} 
                                   & 2700s                          & \multicolumn{1}{c|}{71.2\%}              & 172mAh          & \multicolumn{1}{c|}{69.1\%}              & 193mAh             \\ \cline{2-6} 
                                   & 3000s                          & \multicolumn{1}{c|}{74.8\%}              & 221mAh          & \multicolumn{1}{c|}{71.2\%}              & 289mAh             \\ \hline
\multirow{4}{*}{\textbf{Cifar-10}} & 7500s                          & \multicolumn{1}{c|}{47.1\%}              & 548mAh         & \multicolumn{1}{c|}{45.2\%}              & 657mAh         \\ \cline{2-6} 
                                   & 9000s                          & \multicolumn{1}{c|}{50.7\%}              & 704mAh         & \multicolumn{1}{c|}{48.0\%}              & 884mAh         \\ \cline{2-6} 
                                   & 10500s                         & \multicolumn{1}{c|}{51.9\%}              & 957mAh         & \multicolumn{1}{c|}{48.9\%}              & 1080mAh         \\ \cline{2-6} 
                                   & 12000s                         & \multicolumn{1}{c|}{52.6\%}              & 1190mAh         & \multicolumn{1}{c|}{51.0\%}              & 1394mAh         \\ \hline
\end{tabular}
\label{table:cluster}
\end{table}

\begin{figure}[!tb]
\centering
\subfloat[]{\includegraphics[width=3.5in]{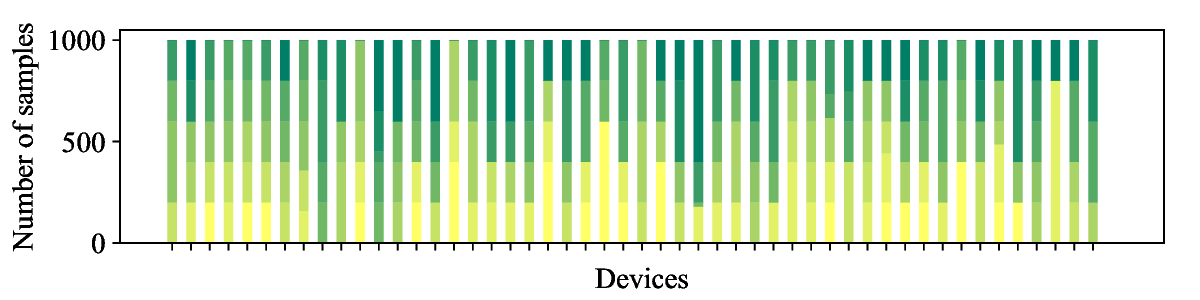}%
\label{fig:noniid}}
\quad
\subfloat[]{\includegraphics[width=3.5in]{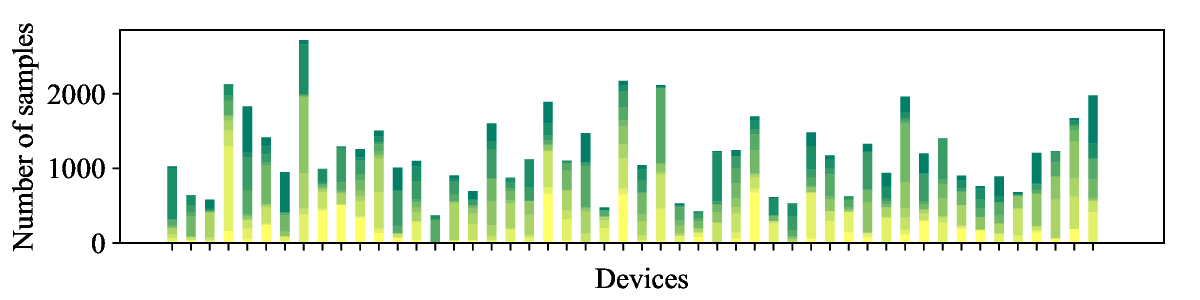}%
\label{fig:direchlet}}
\caption{Different data distributions of the MNIST dataset. (a) The Label non-IID of MNIST (Each device possesses 5 random labels). (b) The Dirichlet non-IID of MNIST (Scaling parameter $\alpha=0.5$).}
\label{fig:noniiddata}
\vspace{-4mm}
\end{figure}

\begin{figure}[!tb]
\centering
\subfloat[]{\includegraphics[width=3.5in]{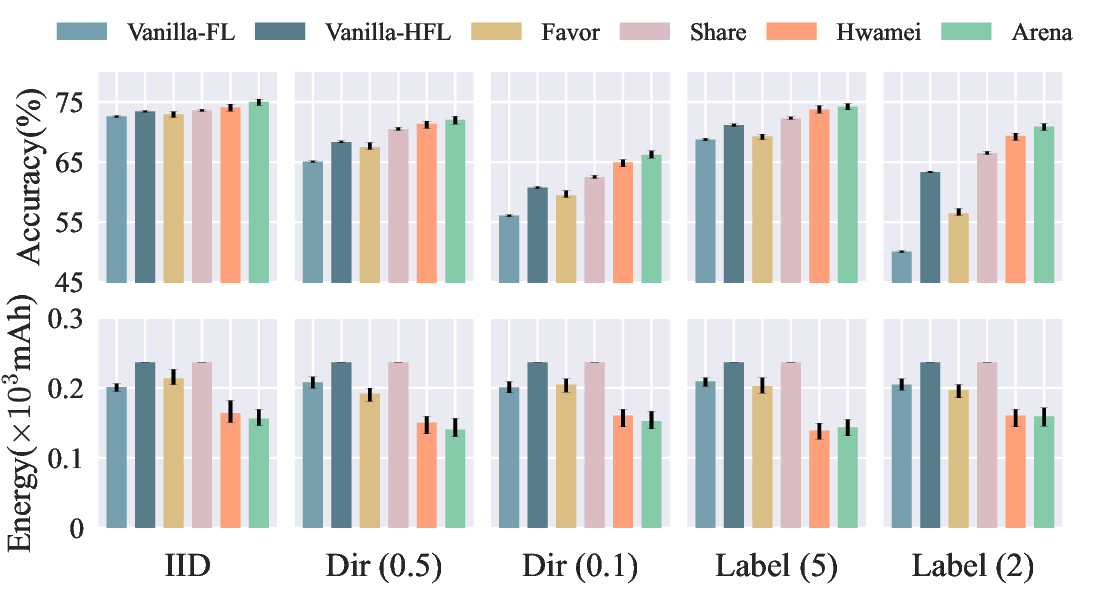}%
\label{fig:result4_MNIST}}
\quad
\subfloat[]{\includegraphics[width=3.5in]{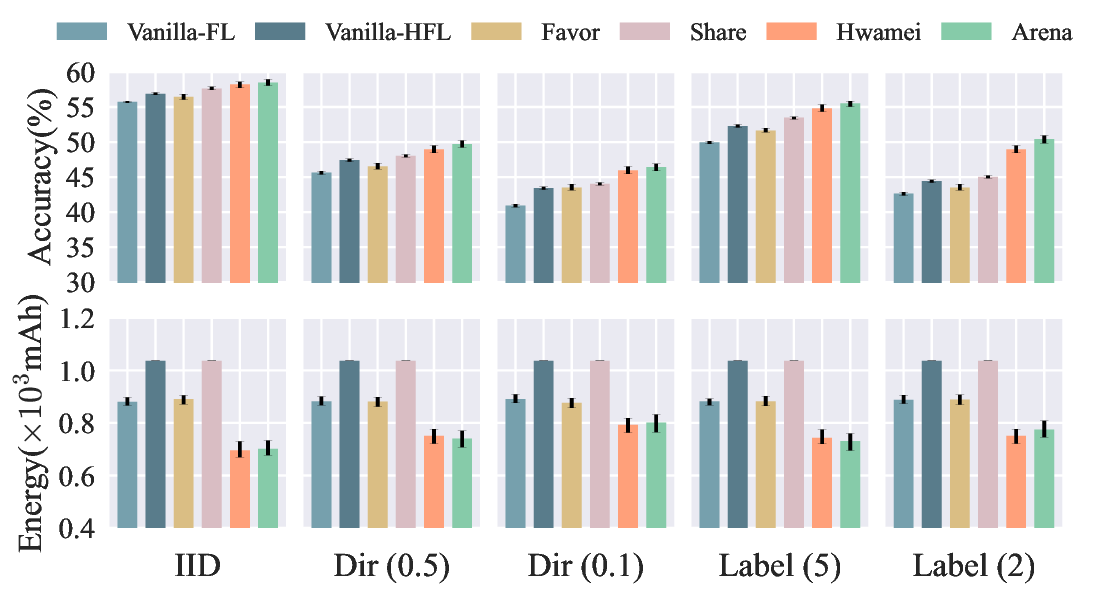}%
\label{fig:result4_Cifar}}
\caption{(a) Accuracy and energy of training MNIST within different non-IID. (b) Accuracy and energy of training Cifar-10 within different non-IID.}
\label{fig:result4}
\vspace{-4mm}
\end{figure}

\subsection{Impact of Profiling Module}

To explore the effect of the clustering component on \textit{Arena}, we compare the performance of \textit{Arena} without profiling module, \textit{i.e.}, keeping the initial topology. 

Table~\ref{table:cluster} shows the impact of clustering on performance. The accuracy we test under the profiling module is generally higher than without clustering. Adding clustering to \textit{Arena} improves accuracy by an average of 3.6\% for MNIST and 4.7\% for Cifar-10 compared to no clustering. This is because clustering eliminates stragglers so that the agent can train the model in the threshold time. We also observe that the energy consumption is generally high without clustering. The profiling module can achieve an average saving of 15.6\% in energy consumption for MNIST and 15.7\% for Cifar-10. In a word, the profiling module enables the system to fully use device resources and produce good performance within a limited time.

\subsection{Different Levels of non-IID Data}

The above experiments are designed to approximate the real data distribution as closely as possible, so each device has only two data labels. In this section, we explore different levels of non-IID data to verify the efﬁciency. We fix the threshold time $T=2500$s for MNIST and $T=8000$s for Cifar-10. For each dataset, we set up three different data distributions, that is IID, Label non-IID and Diriclet non-IID. The different data distribution schemes are shown in Fig.~\ref{fig:noniiddata}.

As shown in Fig.~\ref{fig:result4}, the models’ convergence is sensitive to the level of non-IID of the data distribution. The higher the degree of non-IID, the lower the accuracy of each algorithm. However, \textit{Arena} is the most accurate in different non-IID situations. The improvement of \textit{Arena} is not significant under IID. The higher the degree of non-IID, the more pronounced the advantages of our method, which shows that \textit{Arena} is more robust to real situations with substantial data heterogeneity. Since the time settings are the same, the energy consumption is similar for different distributions. But our algorithm always has the lowest energy consumption of all non-IID levels. In Fig.~\ref{fig:result4_MNIST}, \textit{Arena} saves an average of 24.6\% of the energy per device compared with \textit{Favor} and 34.9\% compared with \textit{Vanilla-HFL} for MNIST. It also saves an average of 29.8\% of the energy per device compared with \textit{Vanilla-HFL} for Cifar-10 in Fig.~\ref{fig:result4_Cifar}. This is because our state design includes data heterogeneity information at edges.

\subsection{Impact of the design of DRL}
\begin{figure}[!tb]
    \centering
    \includegraphics[width=3.2  in]{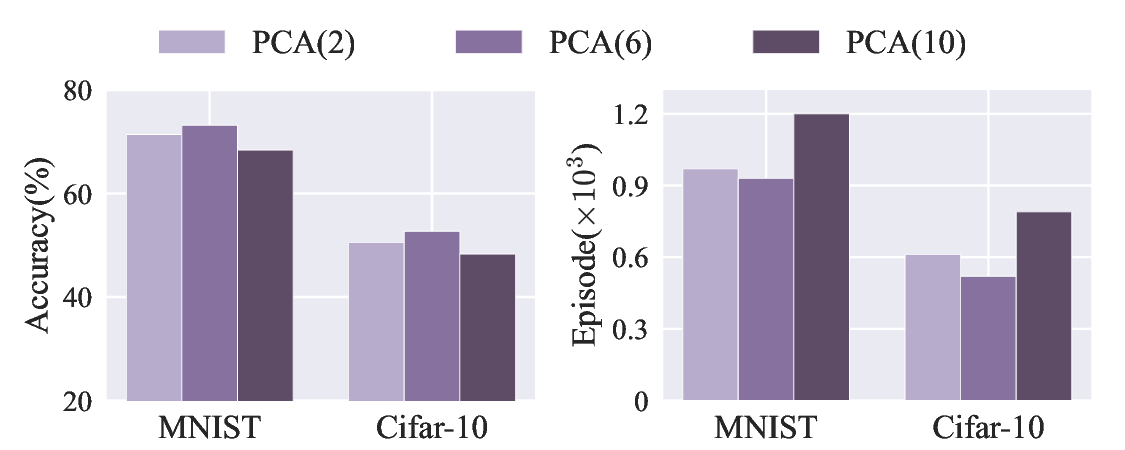}
    \caption{Impact of different principal component}
    \label{fig:pcatest}
    \vspace{-4mm}
\end{figure}

In the design of DRL, the dimensionality of the state can be altered, specifically by modifying the principal component $n_{PCA}$ of the PCA module. We test three different sizes of principal component as shown in Fig.~\ref{fig:pcatest}. We can find in different datasets, the highest accuracy is achieved when the number of principal components is 6, while relatively lower accuracy is observed when it is 2 or 10. Therefore, selecting appropriate PCA parameters is crucial.

Additionally, we analyze the impact of enhancement on DRL. The specific training results are shown in Table~\ref{table:impact}. We test the training accuracy, energy consumption, and convergence episodes of DRL on two datasets. It is easy to observe that, under similar energy consumption, \textit{Arena} converges to the higher accuracy in a shorter number of episodes compared to \textit{Hwamei}.


\begin{table}[]
\caption{Impact of Enhancement}
\centering
\begin{tabular}{cc|c|c|c}
\hline
\multicolumn{2}{c|}{}                                       & \textbf{Accuracy} & \textbf{Energy} & \textbf{Episode} \\ \hline
\multicolumn{1}{c|}{\multirow{2}{*}{\textbf{MNIST}}}    & \textit{Hwamei}     & 72.3\%        & 226mAh      & 1500       \\ \cline{2-5} 
\multicolumn{1}{c|}{}                          & \textit{Arena} & 74.8\%        & 245mAh      & 900       \\ \hline
\multicolumn{1}{c|}{\multirow{2}{*}{\textbf{Cifar-10}}} & \textit{Hwamei}     & 52.1\%        & 1190mAh      & 700       \\ \cline{2-5} 
\multicolumn{1}{c|}{}                          & \textit{Arena} & 52.6\%       & 980mAh     & 500      \\ \hline
\end{tabular}
\label{table:impact}
\end{table}

\section{Related works} \label{Fifth}

As an emerging distribution learning technology, federated learning allows devices to learn models in a distributed system without violating user privacy\cite{FL1, FL2, FL3}. There have been some efforts to improve the federated learning performance in terms of heterogeneity and energy.

System heterogeneity is currently a widely studied issue. 
%
%
Chai \textit{et al.} in \cite{TiFL} divide devices into tiers based on training performance and selecting clients from one tier in each training round. 
%
Stripelis \textit{et al.} in \cite{semi-synchronous} introduce a semi-synchronous FL protocol based on the computational capabilities of devices. 
Cox \textit{et al.} in \cite{Aergia} introduce Aergia that freezing and offloading the models to faster devices. 

The data distribution of devices also influences the statistical heterogeneity\cite{noniid1,noniid2}. Mehryar \textit{et al.} in \cite{Agnostic} introduce an agnostic federated learning framework that avoids biases from non-IID data. 
Wang \textit{et al.} in \cite{Favor} apply DQN to dynamically select devices for training. 
And Pang \textit{et al.} in \cite{DDPG} introduce DDPG to update and manage the collaboration plan among devices. 
Zhang \textit{et al.} in \cite{Dubhe} demonstrate that choosing datasets with similar distributions can accelerate convergence. 
%
%
Nguyen \textit{et al.}\cite{FedDRL} introduce a novel non-IID type and employ an DRL-based algorithm to deal with it. Zhang \textit{et al.}\cite{FedMarl} propose FedMarl based on multi-agent reinforcement learning, utilizing VDN for client selection. 

In addition, it is significant to save energy when deploying the real system. Kim \textit{et al.} in \cite{AutoScale, AutoFL} apply RL to determine the execution targets of devices to maximize energy efficiency. Li \textit{et al.} in \cite{SmartPC} propose a framework for FL, which can balance the training time and model accuracy based on a hierarchical online pace control method. Sun \textit{et al.} in \cite{Energy-aware} propose a device scheduling algorithm that maximizes the choice of devices within the energy constraint. Wang \textit{et al.} in \cite{EEFEI} develop a mathematical model to optimize several vital parameters to minimize energy consumption.

As the scale of the FL systems grows, HFL has been proposed to improve FL's scalability~\cite{HFL, HFLconverge}. 
%
Wang~\textit{et al.} in~\cite{RFLHA} determine the optimal cluster structure and train HFL by a combination of the synchronous and asynchronous algorithms. 
Li~\textit{et al.} in \cite{FedGS} propose FedGS that selects workers by a gradient-based binary permutation algorithm. 
Cui~\textit{et al.} in \cite{FedCS} introduce a heterogeneity-aware heuristic user selection strategy to select devices and determine operating frequency.
%
%
Mhaisen \textit{et al.} in \cite{HFLiid} propose to assign workers to edges so that their data distribution converges to IID.
In contrast, \textit{Arena} is a rare algorithm framework that solves two kinds of heterogeneity and optimizes energy efficiency at the same time. To the best of our knowledge, we are the first to apply DRL to jointly optimize training accuracy and energy consumption based on the HFL framework.

\section{Conclusion} \label{Sixth}

In this paper, we have introduced \textit{Arena}, a learning-based aggregation framework for HFL. Based on pre-experimental results, we have discovered the potential of the HFL framework to improve accuracy and reduce energy consumption by varying the frequencies of cloud aggregation and edge aggregation appropriately. We have designed an DRL-based approach, which collects edge information and global parameters as the state and controls aggregation frequencies to maximize training efficiency. We have built a realistic heterogeneous HFL environment using common devices. Extensive experiments under the realistic framework have been conducted, and the results have demonstrated the efﬁcacy of \textit{Arena} compared with other benchmarks. 

\ifCLASSOPTIONcaptionsoff
  \newpage
\fi



%
%
%
\bibliographystyle{IEEEtran}
\bibliography{cited}

\end{document}